\documentclass[acmsmall,authorversion]{acmart}

\AtBeginDocument{%
  \providecommand\BibTeX{{%
    \normalfont B\kern-0.5em{\scshape i\kern-0.25em b}\kern-0.8em\TeX}}}

\setcopyright{acmlicensed}
\acmJournal{PACMCGIT}
\acmYear{2021} \acmVolume{4} \acmNumber{3} \acmArticle{} \acmMonth{9} \acmPrice{15.00}\acmDOI{10.1145/3480148}

\acmSubmissionID{34}

\usepackage{multirow}
\usepackage{subcaption}
\usepackage[ruled,vlined]{algorithm2e}
\usepackage{wrapfig}

\begin{document}

\title{A GAN-Like Approach for Physics-Based Imitation Learning and Interactive Character Control}

\author{Pei Xu}
\email{peix@clemson.edu}
\author{Ioannis Karamouzas}
\email{ioannis@clemson.edu}
\affiliation{%
  \department{School of Computing}
  \institution{Clemson University}
   \country{USA}
}

\begin{abstract}
We present a simple and intuitive approach for interactive control of physically simulated characters. 
Our work builds upon generative adversarial networks (GAN) and reinforcement learning, and introduces an imitation learning framework where an ensemble of classifiers and an imitation policy are trained in tandem given pre-processed reference clips.
The classifiers are trained to discriminate the reference motion from the motion generated by the imitation policy, while the policy is rewarded for fooling the discriminators. 
Using our GAN-like approach, multiple motor control policies can be trained separately to imitate different behaviors. 
In runtime, our system can respond to external control signal provided by the user and interactively switch between different policies. 
Compared to existing method, our proposed approach has the following attractive properties: 1) achieves state-of-the-art imitation performance without manually designing and fine tuning a reward function; 
2) directly controls the character without having to track any target reference pose explicitly or implicitly through a phase state; and  
3) supports interactive policy switching without requiring any motion generation or motion matching mechanism.
We highlight the applicability of our approach in a range of imitation and interactive control tasks, while also demonstrating its ability to withstand external perturbations as well as to recover balance. 
Overall, our approach has low runtime cost and can be easily integrated into interactive applications and games.
\end{abstract}




\begin{teaserfigure}
\vspace*{.25cm}
    \includegraphics[width=\linewidth]{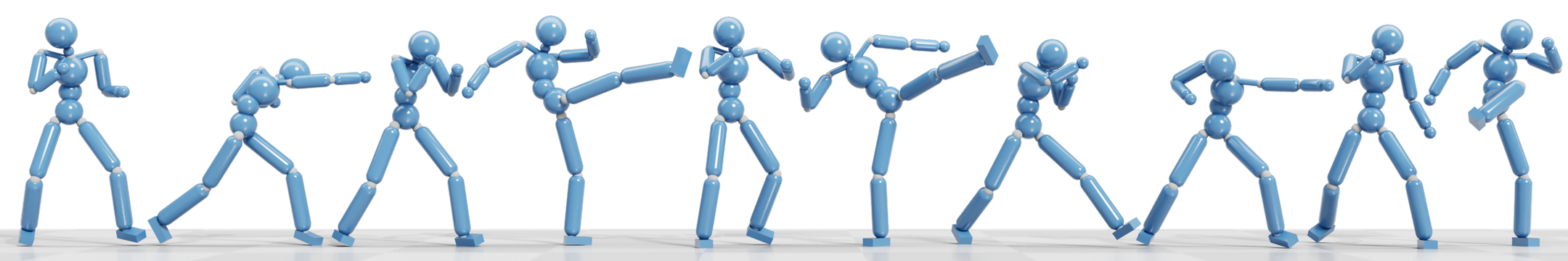}
    \caption{
        A physically simulated character performs diverse punch and kick actions responding to the user's input. 
        Our approach provides a flexible way for real-time interactive control without explicit motion tracking.
    }
    \label{fig:teaser}
\end{teaserfigure}
\maketitle

\section{Introduction}
Data-driven methods have long been used to generate realistic character animation. 
Kinematic approaches typically rely on a pre-collected reference dataset containing diverse motion clips that are used for motion generation or synthesis  
without fully leveraging the physical equations of motion~\cite{holden2017phase,holden2020learned}.
Though high-quality animations can be generated depending on the amount and quality of the motion dataset~\cite{clavet2016motion}, kinematics-based methods may suffer problems when facing complex, unpredicted perturbations or environmental variations.
Physics-based methods, on the other hand, perform simulation through a physics engine to generate motions, and thus guarantee the physical realism of the generated motions~\cite{vbz,wu2010terrain}. 
Such methods can also enable
sim-to-real transfer and apply motion generation techniques on physical robots as shown in recent works~\cite{hwangbo2019learning,RoboImitationPeng20}.

A challenge in physics-based methods for realistic motion generation is that a controller is needed to be optimized 
either directly by applying joint torques or indirectly through, for example, proportional-derivative (PD) servos to help the character reach a desired pose.
However, those control signals are typically inaccessible when we capture motion in the real world.
In recent years, reinforcement learning has been widely used to perform the optimization in order to obtain a general controller without pre-assumed heuristic rules. 
Data-driven methods for imitation learning under the framework of deep reinforcement learning have achieved state-of-the-art performance and are able to generate high-quality motions ~\cite{peng2018deepmimic,nuttapong,ScaDiver}. 
However, when facing an interactive control demand, 
methods that rely on motion tracking usually need a motion generation or matching mechanism 
to ensure the transition between two different behaviors or motor control policies~\cite{bergamin2019drecon,ScaDiver}.

In this paper, we propose an alternative approach for motion capture imitation and interactive control of physically simulated characters based on the framework of generative adversarial imitation learning (GAIL)~\cite{ho2016generative,merel2017learning}.
We propose a number of improvements upon GAIL to stabilize the training process, and 
achieve imitation performance comparable to state-of-the-art methods for motion tracking without the need of designing and fine-tuning a reward function.    
Using our GAN-like approach, multiple motor control policies are trained separately to imitate different reference motion clips.
During runtime, the character can exhibit diverse behaviors through interactive policy switching. 
Our system can run in real time, as checking the feasibility of switching to a target policy is done by simply performing a forward pass on the discriminators 
of the target policy. 
In addition, the policies perform inference using only the last few poses of the character's trajectory without having to track any target pose.
This reduces the system complexity and runtime cost, because no motion generation or matching mechanism is needed for policy transitions.
Even though such a solution requires the reference motions to have some similar poses such that the policy switch can happen, it provides a fast and lightweight alternative to the problem of real-time character control.
We demonstrate the applicability of our approach in a range of imitation and interactive control tasks, while also highlighting the robustness of the trained controllers to external perturbations.

\section{Background and Related Work}

\subsection{Physics-Based Character Control}
Over the past twenty years, various methods have been proposed for physics-based character control in the fields of computer graphics, robotics and control, including
trajectory optimization schemes and reinforcement learning solutions. 
In the related literature on trajectory optimization methods, a number of controllers have been proposed that
focus on specific tasks through manually designed heuristics or feedback rules~\cite{yin2007simbicon,lee2010data,coros2010generalized,brown2013control}. 
Online and offline optimization approaches, including 
model predictive control~\cite{da2008simulation,tassa2012synthesis,tassa2014control}, 
open-loop control~\cite{liu2010sampling,mordatch2012discovery,liu2015improving},
closed-loop feedback control~\cite{mordatch2014combining,liu2016guided,da2017tunable}
and model-based control~\cite{kwon2010control,hamalainen2015online,kwon2017momentum}, 
have also been successfully applied on physics-based character control for motion generation. 
We refer to the survey from~\citet{geijtenbeek2012simple} for a general introduction to a variety of optimization approaches for physics-based character control. 
While characters in most aforementioned works are actuated by joint torques directly or through PD controllers, there are also some interesting works~\cite{geijtenbeek2013flexible,lee2014locomotion, nakada2018deep, lee2019scalable} showing the success in muscle-actuated character control. 

Approaches based on deep reinforcement learning have gained a lot of popularity recently, allowing to train control policies for simulated characters from raw observation data through trial and error.
Goal-conditioned policies that seek to control the character to complete a goal in a physically simulated environment, 
typically without too much concern about the behavior style, have been employed as baselines in many reinforcement learning test suites and recent algorithms~\cite{lillicrap2015continuous,schulman2015high,schulman2015trust,schulman2017proximal,heess2017emergence}.
However, due to the lack of motion style constraint, the generated motions in these works lack the grace and sophistication exhibited by complex beings.
To generate natural motions, many works~\cite{yu2018learning, won2018aerobatics,xie2020allsteps} use carefully designed reward functions or training strategies to encourage the character to act in an expected style.

To obtain realistic motion, lots of recent works on deep reinforcement learning follow a data-driven approach and achieve state-of-the-art performance through learning to imitate reference examples of expert motion. 
DeepLoco~\cite{peng2017deeploco} uses a hierarchical strategy to implement walking style imitation for physics-based humanoid character in navigation tasks. 
DeepMimic~\cite{peng2018deepmimic} enables a physics-based character to exhibit various motion skills learned from artist-authored animation and motion capture data by combining imitation learning with goal-conditioned learning. 
\citet{nuttapong} builds upon DeepMimic to train a single imitation policy from a large collection of clips and a recovery policy that can be used when the character deviates significantly from the reference motion. 
\citet{park2019learning} leverage the kinematic aspects of unorganized motion data to generate reference motions for the character to imitate.
DReCon~\cite{bergamin2019drecon} performs imitation learning with the target pose selected from reference motion dataset dynamically according to the desired movement speed, direction and motion style.
MotionVAE~\cite{ling2020character} employs data-driven generative models using variational autoencoders to generate target motion poses for character control.
ScaDriver~\cite{ScaDiver} proposes a method of learning a policy for diverse behavior control from clustered motion data.

Most imitation learning works mentioned above rely on motion tracking to perform imitation. 
Typically, such methods employ a motion phase state variable~\cite{peng2017deeploco,peng2018deepmimic}, the target pose at next frames~\cite{park2019learning}, or the difference between the character's current pose and the target pose~\cite{ScaDiver,bergamin2019drecon} as a part of the input to the policy network for inference.
In addition, a carefully designed reward function plays a key role for imitation learning in these works.
In this paper, we explore an approach based on generative adversarial networks (GAN)~\cite{goodfellow2014generative}
to perform imitation learning through causal inference based on the character's last pose trajectory only and without any manually designed reward function. 
Concurrent with our work,~\citet{peng2021amp} proposed a GAN-like approach to synthesize high-quality motion from large datasets of unstructured motion clips combined with goal-conditioned learning.
While their work focuses on goal-directed motion synthesis, our approach focuses on interactively controlling physically simulated characters.

\subsection{Generative Adversarial Imitation Learning}
Generative adversarial imitation learning (GAIL)~\cite{ho2016generative} borrows the idea of vanilla generative adversarial networks (GAN)~\cite{goodfellow2014generative} to perform imitation learning.
The vanilla GAN optimizes a generative model to produce real-like data by playing a min-max game with a discriminator trained together with the generator.
Combining the idea of GAN with reinforcement learning, GAIL utilizes a discriminator to identify the agent's actions
from the expert ones, and generates reward signal for reinforcement learning when the agent is able to fool the discriminator. 
GAIL, based on the vanilla GAN, trains the discriminator as a binary classifier to minimize the cross-entropy loss:
\begin{equation}\label{eq:vanilla_gan}
    - E_{(s_t, a_t)}\left[\log D(s_t, a_t)\right] - E_{(\tilde{s}_t, \tilde{a}_t)}\left[\log (1 - D(\tilde{s}_t, \tilde{a}_t))\right],
\end{equation}
where $(s_t, a_t)$ is the state-action pair taken by the agent under training at the time step $t$ and $(\tilde{s}_t, \tilde{a}_t)$ is that taken by the expert.
The work from~\citet{merel2017learning} extends GAIL and proposes to use $(s_t, s_{t+1})$ as a substitute of $(s_t, a_t)$ in the case where the expert action is inaccessible.

GAIL provides a way of imitation learning without interacting with the expert during training.
This is similar to inverse reinforcement learning (IRL)~\cite{russell1998learning, ng2000algorithms} methods,
though IRL explicitly recovers a reward function from expert data before policy extraction.
Through the extension to GAIL proposed by~\citet{merel2017learning}, we can perform imitation learning from motion capture data directly for character control by pose comparison without accessing the control signals.
Despite such recent advancements, GAIL-based methods can be quite unstable because of the difficulty to find the Nash equilibrium between the generator (policy network) and discriminator, resulting in subpar imitation  performance. 

Our approach of imitation learning for physically simulated characters builds off of the GAIL framework from~\citet{merel2017learning}. 
To improve the stability of adversarial training, though, we propose a number of improvements as detailed in Section~\ref{sec:gan}.
Our approach exhibits comparable performance with the state-of-the-art imitation learning methods, and provides a novel way to perform interactive control for physics-based characters.

\section{Overview}
\begin{figure*}[t]
  \centering
  \includegraphics[width=\linewidth]{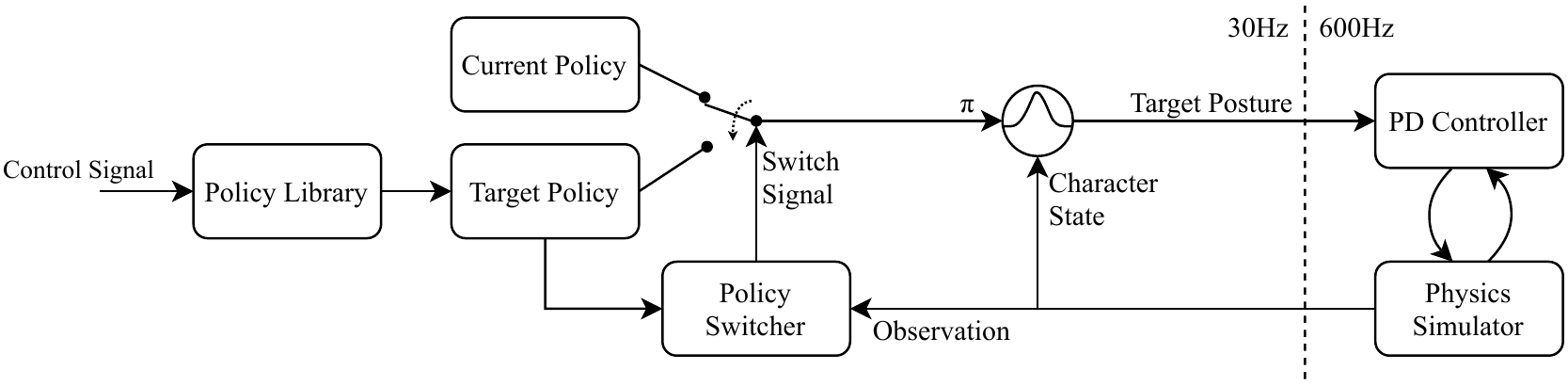}
  \caption{Overview of our system during runtime. 
  The policy switcher uses the character's observation to decide whether to keep the currently activated policy or to switch to the target policy interactively responding to the external control signal 
  when such a transition is considered feasible. 
  The activated policy network queries the character state from the physics simulator and outputs target angles to the PD controller.
  Joint torques are computed from PD servos and applied to the character model.}
  \label{fig:system}
\end{figure*}

Our system implements interactive control of physically simulated characters by
supporting a range of behaviors through imitation learning and policy switching. 
In a pre-processing phase, we first collect a number of motion clips that cover specific behaviors. 
We use these clips to synthesize reference motions that we want the character to imitate (see Section~\ref{sec:results} for details), 
and create a library of physics-based controllers by training corresponding action policies (Section~\ref{sec:control}).   
All policies are trained using a GAN-like approach under the GAIL framework (Section~\ref{sec:gan}). 
As such, we do not have to manually design a reward function for imitation learning. 
We make several improvements over the vanilla GAIL framework and ensure that our GAN-like approach can provide comparable performance to the state-of-the-art imitation learning algorithms for physics-based characters.

During runtime, the user can select a target behavior to be performed by the character via any input device. 
The system responds to the user's control signal and attempts to switch the current policy to the target one,  
which enables the character to perform a specific action by providing a target posture to angular PD servos. 
Policies trained with our approach imitate given reference motions and perform inference based only on the short-term pose trajectory of the character. 
Therefore, they can directly take over the character when given a pose trajectory similar to that in the reference motions, without having to track any target reference pose.
We exploit the discriminators trained with the target policy as a \emph{policy switcher} 
to measure the similarity of the current pose trajectory of the character to the reference motions,
and decide if the transition to the target policy is feasible.
Figure~\ref{fig:system} shows a schematic overview of our system during runtime. 

\section{GAN-like Approach for Imitation Learning}\label{sec:gan}
The vanilla GAIL framework~\cite{ho2016generative}  performs reinforcement learning by generating reward signals through a discriminator,
which outputs a probabilistic result denoting the likelihood of a given state-action sample to be drawn from the expert policy. 
However, this method cannot be directly applied to the problem of  imitation learning from motion clips, 
as, in general, we cannot directly access the actions (joint torques or target postures to PD servos).
As such, we follow the strategy from~\citet{merel2017learning} and use transitions of the character's observations as the input to the discriminator.

\begin{wrapfigure}{r}{0.6\textwidth}
  \vspace{-0.4cm}
  \includegraphics[width=\linewidth]{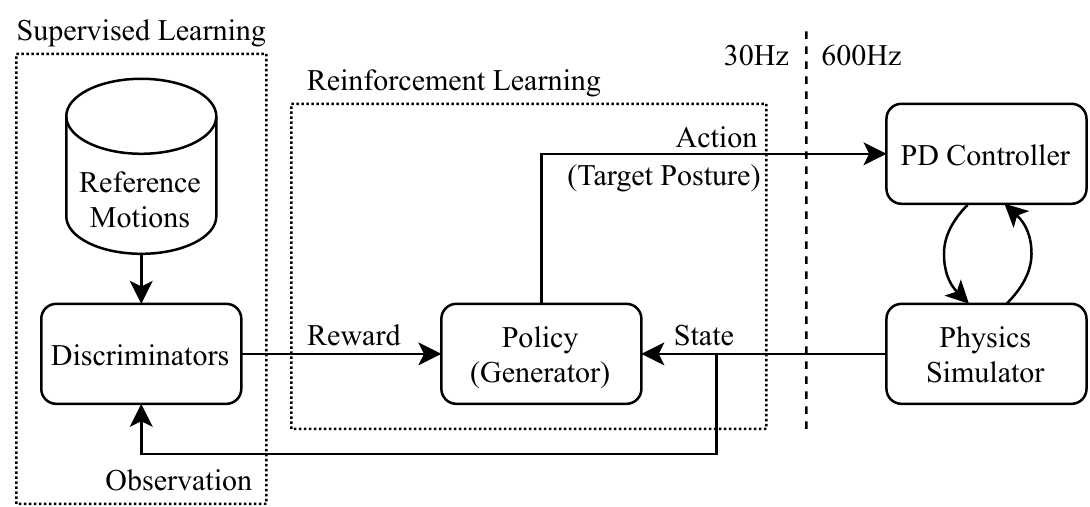}
  \caption{
  Schematic overview of training a physics-based character controller from reference motion using our GAN-like approach with reinforcement learning.
  }
  \label{fig:gan}
\end{wrapfigure}
A typical problem of using a GAN-like approach for reinforcement learning is reward signal vanishing.
In such a case, the discriminator is too harsh and always gives a very low reward to the agent under training regardless of what progress the agent makes, which result in the training being unable to carry on. 
This issue is actually similar to the problem of gradient vanishing in vanilla GAN applications, 
often caused by over-optimization of the discriminator.
An exactly opposite situation occurs when the discriminator is under-optimized and unable to effectively identify observation samples from the agent under training. 
As a result, the agents under such a situation would be ``self-satisfied'' and cannot be improved to learn to imitate the reference motion accurately.
To solve these problems and improve the stability and performance for imitation learning from motion clips through a GAN-like approach, we adopt three tricks: 
1) leveraging multiple discriminators as an ensemble; 
2) using hinge loss for discriminator training instead of the binary cross entropy loss employed by GAIL; and 
3) introducing a gradient penalty term~\cite{gulrajani2017improved} during discriminator optimization. 
We refer to Fig.~\ref{fig:gan} for a schematic overview of the training process of our proposed GAN-like approach.  

\subsection{Observation Space}\label{sec:ob_space}
We use the local position and orientation 
(measured in the unit of quaternion) of the character links relative to the root position and heading direction as the observation space of the discriminator. 
To embed the movement of the character in the simulated world, 
we take into account the observations from five consecutive frames relative to the root position and heading direction in the last frame.
Given the observation $\mathbf{o}_{t-3:t+1} \in \mathbb{R}^{5\times7\vert\mathcal{B}\vert}$
where $\mathcal{B}$ is the collection of the character body links, 
the $i$-th discriminator, $D_i(\mathbf{o}_{t-3:t+1})$, 
is trained to identify if the pose trajectory represented by $\mathbf{o}_{t-3:t+1}$ is from the reference motion or is taken by the agent under training.
This provides the appropriate reward signal at time step $t$ for policy training using reinforcement learning. 

\subsection{Multiple Discriminators}
An ignorant agent trained from scratch typically needs to do a large amount of exploration early on during training before reaching some poses similar to those in the reference motion. 
In contrast, an effective discriminator could easily identify samples from the character agent at the early phases of training and generate low reward signals,
which do not encourage the agent's exploration. 
To address this issue,
We employ an ensemble discriminator using the average score from multiple discriminators.
Ensemble is a common machine learning technique to produce better predictive model based on multiple base models~\cite{dietterich2000ensemble}. 
Work from~\citet{durugkar2016generative} explores the feasibility of using GAN with an ensemble discriminator in image generation tasks.

In our implementation, multiple discriminators are initialized randomly but share the same feature extraction layers.
Such an architecture can help avoid model collapse caused by overfitting of single discriminator.
It could also prevent reward signal vanishing at the early stages of training by increasing the difficulty of optimization caused by the discrepancy of multiple discriminators. 
In Section~\ref{sec:results}, we empirically show 
the effect that different number of discriminators have on the performance.

\subsection{Hinge-Loss Optimization}
To avoid over-optimization of the discriminator, 
we use a hinge version of the discriminator loss (Eq.~\ref{eq:vanilla_gan}) as proposed by~\citet{lim2017geometric}:
\begin{equation}
\label{eq:hinge}
    \mathcal{L}_{D_i} = \mathbb{E}\left[\max(0, 1+D_i(\mathbf{o}_{t-3:t+1}))\right] + \mathbb{E}\left[\max(0, 1-D_i(\mathbf{\tilde{o}}_{t-3:t+1}))\right],
\end{equation}
where $\mathbf{o}_{t-3:t+1}$ are observation samples from the character agent and $\mathbf{\tilde{o}}_{t-3:t+1}$ are samples drawn from reference motion clips that the character agent intends to imitate.
This loss function borrows the idea of support vector machine optimization with soft margin.
Minimizing Eq.~\ref{eq:hinge} 
is equivalent to optimizing the discriminator to perform classification based on the separating hyperplane of 0 with margin of 1.
This loss function prevents over-optimization by avoiding performing optimization over samples that have been classified correctly, i.e. samples satisfying $D_i(\mathbf{o}_{t-3:t+1}) \leq -1$ or  $D_i(\mathbf{\tilde{o}}_{t-3:t+1}) \geq 1$.
GAN with hinge loss has been adopted by many works in the field of image generation~\cite{brock2018large, miyato2018spectral, zhang2019self}.

When switching to a hinge loss, the discriminators do not provide anymore a probabilistic result, 
but a score, with a valid range of [-1, 1], expressed as a \textit{monotonic} measurement of the similarity between a given pose trajectory and segments 
in the reference motion. 
This property is important, as we can use this score as a reward signal that has a clearly defined upper and lower bound, and monotonically increases with the similarity of the given pose trajectory and the reference motion.

\subsection{Gradient Penalty}
The gradient penalty term was introduced by WGAN-GP~\cite{gulrajani2017improved} as a regularization term added to discriminator loss function to enforce the 1-Lipschitz constraint, thereby preventing gradient explosion and vanishing, and improving the stability of training.
The gradient penalty term for the $i$-th discriminator 
in our implementation is computed as
\begin{equation}
    \mathcal{L}_{D_i}^{GP} = \mathbb{E}\left[(\vert\vert \nabla_{\mathbf{\hat{o}}_{t-3:t+1}} D_i(\mathbf{\hat{o}}_{t-3:t+1}) \vert\vert_2 - 1)^2\right],
\end{equation}
where $\mathbf{\hat{o}}_{t-3:t+1} = \alpha \mathbf{\tilde{o}}_{t-3:t+1} + (1-\alpha) \mathbf{o}_{t-3:t+1}$ with the interpolation coefficient $\alpha$ drawn uniformly between 0 and 1
for each training sample pair ($\mathbf{\tilde{o}}_{t-3:t+1},\mathbf{o}_{t-3:t+1}$).  
$\mathbf{\hat{o}}_{t-3:t+1}$ denotes the interpolation of samples between the reference motion and the ones from the agent under training,
and is exploited to emulate all possible samples distributed along the mainfold of the valid observation space. 
This regularization term actually intends to optimize the discriminator network such that the gradient norm with respect to all possible observations is close to 1. 

Overall, 
our ensemble with $N$ discriminators approach has the following loss function: 
\begin{equation}\label{eq:loss_d}
    \mathcal{L}_D = \frac{1}{N} \sum_{i=1}^N \left(\mathcal{L}_{D_{i}} + \lambda^{GP} \mathcal{L}_{D_i}^{GP}\right),
\end{equation}
where $\lambda^{GP}$ is a scalar coefficient.
In our implementation, all discriminators share the same layers before the final one, as shown in Fig.~\ref{fig:network}.
The final layer linked to each discriminator is initialized independently and randomly to provide  diversity. 
Our approach of discriminator ensemble is different than average pooling. 
As the score, $\mathcal{L}_{D_{i}}$, of each discriminator is computed independently before applying the average operation, the hinge loss in Eq.~\ref{eq:hinge} will prevent optimization through those discriminators that perform classification correctly, and refrain individual discriminators from contributing too large or small scores. 
In fact, applying an extra average pooling layer to the final layer of the discriminator network is equivalent to 
converting the final fully connected layer to having 1-dimensional output, 
i.e. the architecture of one discriminator without any ensemble,
which is also susceptible to overfitting (see Section~\ref{sec:sensitivity}).

\section{Dynamic Controller Learning}\label{sec:control}
Following the GAIL framework, we train action policies as dynamic controllers via reinforcement learning without having to  manually design a reward function. Instead,  we exploit the discriminator output as the reward signal. 

\subsection{State and Action Space}\label{sec:state_space}
State-of-the-art imitation learning implementations for physics-based character control typically construct the state space by leveraging the target pose implicitly through a phase variable~\cite{holden2017phase,peng2018deepmimic} or explicitly~\cite{park2019learning,bergamin2019drecon,ScaDiver} by using the target pose directly or the deviation between the target and the current pose of the character.
When employing such a kind of action policies for interactive control, the system needs to be able to generate a target pose~\cite{lee2019scalable} or find a proper one from 
the reference motion dataset~\cite{clavet2016motion,holden2020learned}.
Such a process, though, could be time consuming. 
In contrast, in our framework, the policy network performs target pose inference based on a small number of previously encountered frames (similar to the observation space outlined in Section~\ref{sec:ob_space}),
and directly outputs target postures that are given as input to PD servo controllers.

As opposed to the discriminator, action policies need to generate reasonable target posture depending not just on the character pose but also on the current velocity.
Therefore, we augment the observation space and set the state space for action policies as
$\mathbf{s}_t := \mathbf{o}_{t-3:t}^{+v} \in \mathbb{R}^{4\times13\vert\mathcal{B}\vert}$,
where $\mathbf{o}_{t-3:t}^{+v}$ is the augmented observation based on four consecutive frames from $t-3$ to $t$ (inclusive) containing also the velocity state of character body links relative to the root heading direction at the $t$-th frame.
We use a recurrent neural network (RNN),
a GRU~\cite{chung2014empirical} layer in our implementation, to embed the input state defined in the temporal space.

The action space is the target posture for each joint and serves as 
the control signal fed to the PD servos. 
For revolute joints, the 1-dimensional target posture denotes the desired angle position in radians.
For spherical joints, the target posture is a 4-dimensional rotation expressed in the format of axis and angle.
The action policy is defined by a normal distribution with dependent components, the mean value of which and the logarithm of the standard deviation are obtained from the output of the policy network.

\subsection{Behavior Learning}\label{sec:diverse_behavior_policy}
We combine our discriminator learning approach with a policy gradient algorithm for reinforcement learning to find an imitation policy $\pi$ that maximizes the cumulative reward  $\sum_t \gamma^{t-1} r_t(s_t, a_t, s_{t+1})$ of collected experiences where $\gamma$ denotes the discount factor. 
We compute the reward, $r_t$, for each state-action-state transition by considering the average score of the ensemble of $N$ discriminators. 
The score from each discriminator is clipped before averaging according to hinge loss definition in Eq.~\ref{eq:hinge}, which avoids biasing the reward towards too low or small scores from individual discriminators:
\begin{equation}\label{eq:reward}
    r_t(\mathbf{s}_{t}, \mathbf{a}_t, \mathbf{s}_{t+1}) = \frac{1}{N} \sum_{i=1}^N clip(D_i(\mathbf{o}_{t-3:t+1}), -1, 1).
\end{equation}
Given the state-observation pair $\{(\mathbf{s}_{t}, \mathbf{s}_{t+1}); \mathbf{o}_{t-3:t+1}\}$,
this reward evaluates 
the agent's performance to reach a desired target pose at frame $t+1$
given the character's trajectory from frame $t-3$ to $t$. 

During training, by interpolating the keyframes, a pose randomly drawn from the reference motion is used to initialize the character model in each episode. 
The interpolation prevents the discriminators from simply memorizing the poses in the keyframes, and helps the policy for character control to generalize better.
After pose initialization, at each time step $t$, the agent receives the reward signal $r_t$ from the discriminator ensemble.
An episode ends if the time step reaches the maximal length of the reference motion clip for a non-cyclic motion, or if an overtime limit, 500 frames in our implementation, is hit for a cyclic motion.
An early termination is taken if an unexpected contact between the character body link and the ground occurs, e.g,. if the character falls down and gets in contact with the ground with its chest link when imitating the walk motion. 
The policy is updated, through a reinforcement learning algorithm, which is DPPO~\cite{heess2017emergence} in our implementation,  after a batch of state-action-reward samples is collected.
The ensemble discriminator is trained together with the policy  using observation samples collected from the agent under training
and the same amount of samples drawn from the reference motion.
Algorithm~\ref{fig:alg} summarizes the training process using our GAN-like approach.

\begin{algorithm}[t]
\SetAlgoLined\small
 Initialize action policy network $\pi$\;
 initialize policy replay buffer $\mathcal{T}$\;
 initialize discriminators network $D_i$ where $i = 1, \cdots, N$\;
 initialize discriminator replay buffer $\mathcal{M}$\;
 preprocess reference motion dataset $\mathcal{K}$\;
 initialize simulation environment.
 
 \While{training does not converge}{
      \For{each environment step $t$}{
            $\mathbf{a}_t \sim \pi(\mathbf{s}_t)$ \;
            $\mathbf{s}_{t+1}, \mathbf{o}_{t-3:t+1} \gets$ environment updates with character control signal of $\mathbf{a}_t$\;
            compute $r_t$ using  Eq.~\ref{eq:reward}\;
            draw the sample $\mathbf{\tilde{o}}_{t-3:t+1}$ from $\mathcal{K}$\;
            $\mathcal{M} \gets \mathcal{M} \cup \{\mathbf{o}_{t-3:t+1}, \mathbf{\tilde{o}}_{t-3:t+1}\}$\;
            $\mathcal{T} \gets \mathcal{T} \cup \{(\mathbf{s}_t, \mathbf{a}_t, r_t, \mathbf{s}_{t+1})\}$\;
            $\mathbf{s}_t \gets \mathbf{s}_{t+1}$.
      }
      
      \For{each discriminator update step}{
            update ensemble discriminator $D$ based on Eq.~\ref{eq:loss_d} using samples from $\mathcal{M}$.
      }
      $\mathcal{M} \gets \emptyset$.
      
      \For{each policy update step}{
            update $\pi$ using samples from $\mathcal{T}$.
      }
 }
 \caption{GAN-like approach for imitation learning}
 \label{fig:alg}
\end{algorithm}

Our approach can be easily extended for 
\emph{integrated learning} of a single policy from multiple reference motion clips, i.e., training one policy capable of imitating pose trajectories from a collection of multiple reference clips.
To do so, at the beginning of each training episode, we randomly choose one of the clips from the collection of target, reference clips for imitation.
Such an approach can simplify the interactive system for character control, 
which we will introduce in Section~\ref{sec:policy_switch},
as we can use a single policy to enable the character to automatically perform different behaviors  
according to its current pose when a policy switch is triggered.

\subsection{Interactive Policy Switch}\label{sec:policy_switch}
At runtime, there is only one action policy activated that controls the character by outputting target postures to PD servos. 
Our system allows for interactive character control by letting the user specify a target action policy that can take over the character and perform a specific behavior. 
If the target policy can imitate its corresponding reference motion well, when the current pose of the character is similar to some pose in the reference motion, the target policy should  be able to take over the character and perform the behavior shown in the reference motion.
Based on this assumption, by exploiting the ensemble of discriminators  trained in tandem with the policy network, 
we use the discriminator score to implicitly estimate the similarity of the current pose trajectory $\mathbf{o}_{t-4:t}$ to segments in the reference motion of the target policy,
and consider the switch to the target policy to be feasible if
\begin{equation}\label{eq:switch_cond}
    \frac{1}{N} \sum_{i=1}^N clip(D_i^{target}(\mathbf{o}_{t-4:t}), -1, 1) \geq \tau,
\end{equation}
where $D_i^{target}$ is the $i$-th discriminator of the target policy and $\tau$ is a threshold scalar value.

As detailed in Section~\ref{sec:state_space}, 
policy networks perform imitation by causal inference based only on the character's last pose trajectory.
Therefore, when performing policy switch, we do not need to explicitly generate a target pose or track the reference motion by keyframes or a phase state variable, which decreases the  
runtime cost of the interactive system. 
In addition, the transitioning between poses from different behaviors can happen at the intermediate states between keyframes,
as the discriminators are trained using interpolated keyframe samples from the reference motion instead of simply memorizing the keyframes.
To improve the robustness of policy switch between motions having similar but not exactly the same poses,
we add some noise to the character's initial pose at each training episode and also to the corresponding samples drawn from the reference motion for discriminator learning.

After receiving a control signal, our system keeps checking the switch condition shown in Eq.~\ref{eq:switch_cond} at each frame,
and sets the target policy as the current policy if the condition is met
or ignores the control signal if the condition cannot be satisfied within an acceptable response time. 
Our approach only relies on pre-trained policies to control the character performing specific behaviors,
and does not generate any transient states during the running time of policy switch.
Therefore, to expect a transition from one motion to another to be feasible during runtime, system designers must ensure that the two motions have similar poses.
Motion matching~\cite{clavet2016motion, bergamin2019drecon, holden2020learned} technique can be employed as a method to perform similarity check between two motion clips during the data preprocessing phase.

\section{Experiments}
\label{sec:results}
In this section, we experimentally evaluate our work focusing on two aspects: 1) the motion imitation ability of policies trained by our GAN-like approach; and 2) the effectiveness of the implemented interactive control system that exploits the trained policies. 
We also perform sensitivity analysis on the learning performance using different discriminator training strategies. 

\subsection{Setup}\label{sec:setup}
\begin{figure*}[t]
\begin{minipage}[t]{0.4\linewidth}\vspace{0pt}
\footnotesize
  \begin{tabular}{lc}
    \toprule
    Parameter & Value\\
    \midrule
    policy network learning rate & $5 \times 10^{-6}$\\
    value network learning rate & $1 \times 10^{-4}$\\
    discriminator learning rate & $1 \times 10^{-5}$\\
    reward discount factor ($\gamma$) & $0.95$ \\
    GAE discount factor ($\lambda$) & $0.95$ \\
    surrogate clip range ($\epsilon$) & $0.2$ \\
    gradient penalty coefficient ($\lambda^{GP}$) & $10$ \\
    PPO replay buffer size & $4096$ \\
    PPO batch size & $256$ \\
    PPO optimization epochs & $5$ \\
    discriminator replay buffer size & $8192$ \\
    discriminator batch size & $512$ \\
  \bottomrule
\end{tabular}
  \captionof{table}{Hyperparameters}
  \label{tab:hyper}
\end{minipage}
\hfill
\begin{minipage}[t]{0.55\linewidth}\vspace{8pt}
  \centering
  \includegraphics[width=\linewidth]{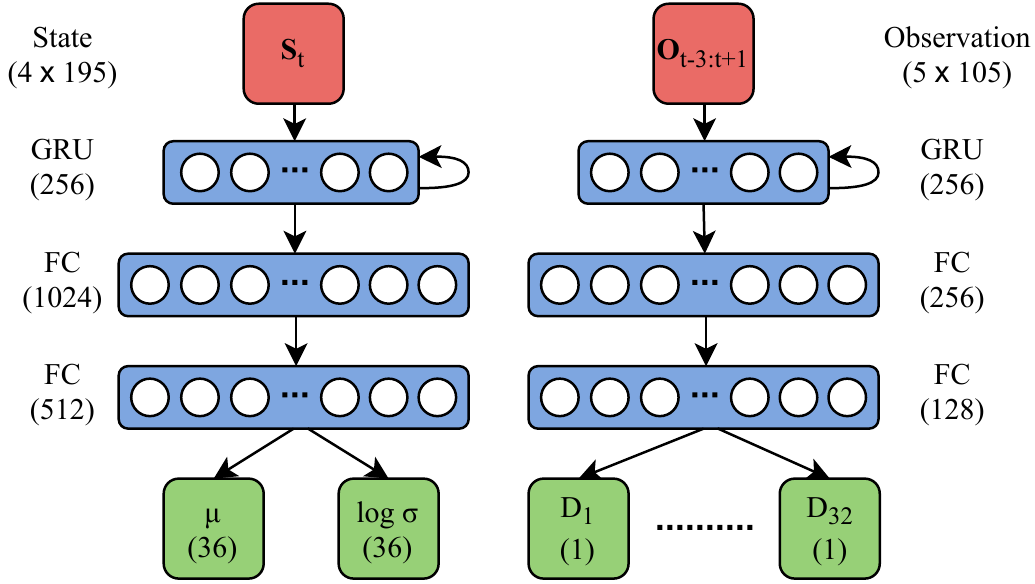}
  \caption{Policy and discriminator network architecture.}
  \label{fig:network}
\end{minipage}
\end{figure*}

We set up our simulation environment using PyBullet~\cite{coumans2021}. 
In all our experiments, we use a humanoid character that is 1.62m tall and weights 45kg. 
The simulated character has 8 spherical joints and 4 revolute joints, plus 2 end-effectors (hands) connected to the forearms with fixed joints and a root link, resulting in 15 body links with 34 degrees of freedom in total.
The character is controlled through stable proportional derivative controllers~\cite{tan2011stable}. 
The whole system interacts with user input and queries character control signals from the policy network at 30 Hz,
while running forward dynamics simulation at 600 Hz. 
We model the control signal for each spherical joint to the PD servos as a quaternion resulting in a 36-dimensional action space.
As described in Sections~\ref{sec:ob_space} and~\ref{sec:state_space}, 
the policy network takes as input four consecutive frames with state space of $\mathbb{R}^{4\times195}$,
while the observation space for discriminators is $\mathbb{R}^{5\times105}$.

In our experiments, we employ 8 threads to perform distributed training under the framework of DPPO~\cite{heess2017emergence}, 
and use Adam optimizer~\cite{kingma2014adam} to perform policy optimization.
DPPO uses an actor-critic architecture to perform reinforcement learning.
The policy (actor) network has one GRU layer with 256 hidden neurons to embed the temporal state input, followed by 2 fully-connected (FC) layers with hidden neurons of 1024 and 512 respectively.
We use a multivariate Gaussian distribution with independent components as the policy distribution. The policy network outputs the mean and standard deviation parameters of the policy distribution. 
The value (critic) network is trained to estimate the state value function, and
has the same architecture with 1-dimensional output.
We setup multiple discriminators sharing the same feature extraction layers, which have similar architecture with the policy network but use fully-connected layers with 256 and 128 neurons following the GRU layer.
By default, we use an ensemble of 32 discriminators initialized randomly.
Discriminator weights are initialized by orthogonal initializer~\cite{saxe2013exact}. The policy network is initialized by truncated normal distribution with a small value of standard deviation 0.05 to prevent too large output at the early training.
Input state and observation to the networks are normalized by a moving average that is dynamically updated during training.
All the hyperparameters and network architecture are shown in Table~\ref{tab:hyper} and Fig.~\ref{fig:network}, respectively.

\subsection{Data Acquisition}\label{sec:data_acq}
All reference motion clips used in our experiments are extracted from the LAFAN1 dataset~\cite{harvey2020robust}.
The motion data provided by LAFAN1 are captured from different subjects at 30Hz.
Each data item contains a motion capture recording, typically having a length of 3-5 minutes, 
with varieties of behaviors under a common motion category performed by one subject. 
We synthesize a short reference motion clip (1-3s) by extracting a motion segment of an interesting single behavior from consecutive frames recorded in a data item,
and retarget the local joint position from the motion data to our character model without any manual reprocessing.
We demonstrate that our approach can work well with raw motion capture data, even when the character model does not match perfectly the recorded subjects.
Table~\ref{tab:motions} gives a description of the behaviors contained in the reference motion clips used in our experiments.
We refer to the supplementary document for details of all the reference motion clips that we used for policy training during experiments.

\subsection{Motion Clip Imitation}\label{sec:res_imit}
\begin{figure}[t]
  \centering
  \begin{subfigure}[b]{0.495\linewidth}
  \includegraphics[width=\linewidth]{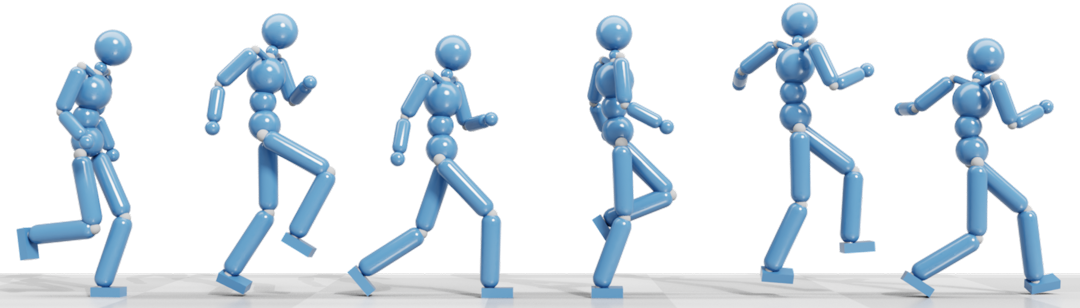}
  \caption{Jaunty Walk}
  \end{subfigure}\hfill
  \begin{subfigure}[b]{0.495\linewidth}
  \includegraphics[width=\linewidth]{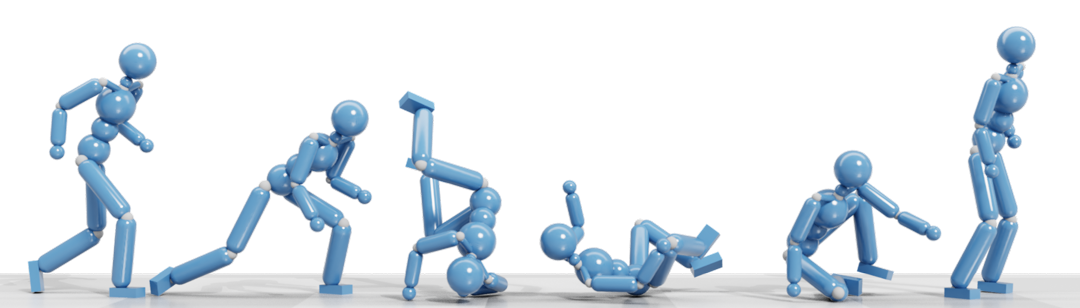}
  \caption{Roll}
  \end{subfigure}

  \begin{subfigure}[b]{0.495\linewidth}
  \includegraphics[width=\linewidth]{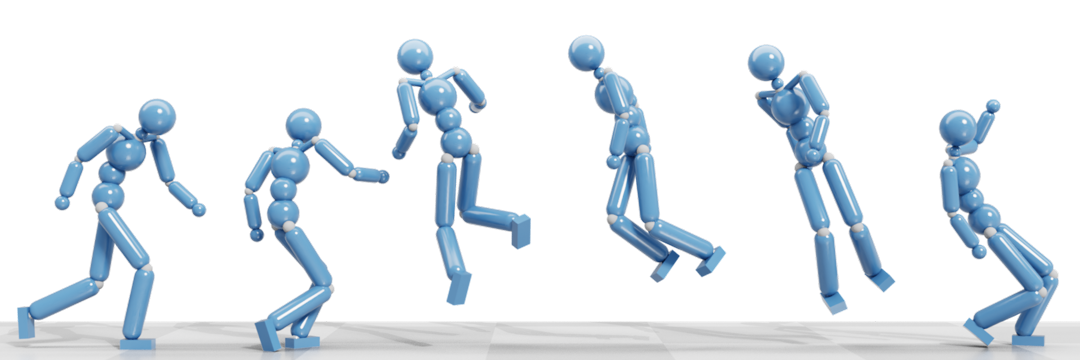}
  \caption{Spinning Jump}
  \end{subfigure}\hfill
  \begin{subfigure}[b]{0.495\linewidth}
  \includegraphics[width=\linewidth]{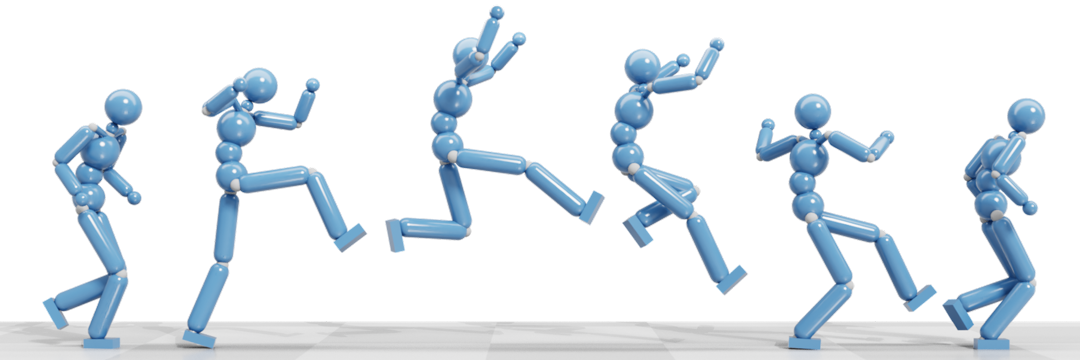}
  \caption{Long Jump}
  \end{subfigure}
  
  
  \caption{Motion poses captured from the policies trained by our GAN-like approach.
    Table~\ref{tab:motions} gives a description of these motions. We refer to the supplemental video for more examples.
  }
  \label{fig:motions}
\end{figure}

\begin{table*}
  \footnotesize
  \begin{tabular}{lccl}
    \toprule
    Motion & Length [s] & Imitation Error [m] & Description\\
    \midrule
    walk & 1.10 & $0.06 \pm 0.02$ & normal walking with a speed around 1.2m/s\\
    pace & 1.70 & $0.11 \pm 0.04$ & slow walking with arms akimbo\\
    limp & 1.90 & $0.09 \pm 0.06$ & slow walking with right leg hurt\\
    swaggering walk & 1.07 & $0.05 \pm 0.01$ & exaggerated walking with one arm akimbo\\
    sashay walk & 1.07 & $0.04 \pm 0.02$ & walking in a slightly exaggerated manner\\
    jaunty walk & 1.40 & $0.15 \pm 0.07$ & walking in a spirited manner \\
    stomp walk & 1.23 & $0.05 \pm 0.04$ & walking while stomping on the ground\\
    stoop walk & 0.93 & $0.05 \pm 0.02$ & slow walking with body bent over\\
    joyful walk & 1.20 & $0.08 \pm 0.03$ & strut walking rhythmically\\
    walk with arms akimbo & 2.20 & $0.10 \pm 0.08$ & high knee walking with arms akimbo\\
    sharp turn during running & 1.47  & $0.30 \pm 0.05$ & 90-degree sharp turning during running\\
    90-degree turn during walking & 2.07 & $0.10 \pm 0.04$ & 90-degree turning during a normal walk\\
    roll & 3.37 & $0.26 \pm 0.06$ & forward rolling from a standing pose\\
    fight stance & 0.87 & $0.05\pm0.01$ & standing with a fighting pose\\
    run & 2.37 & $0.10 \pm 0.04$ & running starting from a walking pose\\
    spinning jump & 3.00 & $0.16 \pm 0.04$ & jumping and spinning the body in the air\\
    long jump &  1.77 & $0.21 \pm 0.07$ & jumping after a running approach\\
    get up & 2.87 & $0.21 \pm 0.03$  & getting up from stumbling \\
    punch & 1.73 & $0.09 \pm 0.06$ & left and then right straight punch \\
    kick &  2.33 & $0.14 \pm 0.06$ & side kick \\
  \bottomrule
\end{tabular}
\caption{Imitation performance when learning from a single motion clip.
Reported numbers denote mean values over 20 trials $\pm$ standard deviation.}
  \label{tab:motions}
\end{table*}

\begin{table}
\footnotesize
  \begin{tabular}{lcccp{5cm}}
    \toprule
    Motion Collection & \# of clips & Length [s] & Imitation Error [m] & Description\\
    \midrule
    run & 2 & 4.73 & $0.15 \pm 0.05$ & running starting from walking poses with different standing foot \\
    spinning and long jump & 4 & 9.53 & $0.37 \pm 0.19$ & spinning jump and long jump with different take-off foot \\
    get up & 5 & 13.7 & $0.18 \pm 0.06$ & getting up from stumbling with different poses lying on the ground\\
    punch & 3 & 4.77 & $0.12 \pm 0.10$ & punches with different stance \\
    kick & 4 & 8.43 & $0.13 \pm 0.11$ & kicks with different stance \\
  \bottomrule
\end{tabular}
 \caption{Imitation performance of policies learning from multiple reference clips in a collection.  
Each collection has different behaviors under the same category of motions. 
A single policy is trained in an integrated manner to learn all the behaviors in one collection.
Results are the evaluation over 20 trials of each clip that a policy learns to imitate.
}
  \label{tab:motion_collections}
\end{table}

We show motion pose snapshots captured from some of our trained imitation policies in Fig.~\ref{fig:motions}.
All policies were trained on a single machine with Intel 6148G CPU and Nvidia V100 GPU. 
It takes about 12 hours to train a policy using a fixed budget of 20 million samples (environment steps).

To quantitatively evaluate the imitation ability of our GAN-like approach, we measure the imitation error as the average global position error of character body links compared to the reference motion:
\begin{equation}
    e = \frac{1}{\vert \mathcal{B} \vert} \sum_{i\in\mathcal{B}} \vert\vert p_{i,t} - p_{i,t}^{ref}\vert\vert_2,
\end{equation}
where $\mathcal{B}$ is the collection of character body links, $p_{i,t}$ is the 3D global position of the $i$-th link at frame $t$ and $p_{i,t}^{ref}$ is the position of the corresponding link in the reference motion.
Since our approach does not implicitly employ any target pose or explicitly exploit any time component or phase state to perform synchronization with the reference motion, a small pose error caused by desynchronization occurring at early frames may result in a large accumulated error even if the subsequent motion is performed smoothly. 
To synchronize the motion gaits, we compute the error of one motion cycle and perform multiple test trials starting at different initial poses.
Table~\ref{tab:motions} lists all the reference motion clips used in our experiments for single-clip imitation tests and the corresponding imitation errors obtained with our approach.
Our approach can imitate the reference motion closely without requiring any explicit reward function fine-tuning.

\begin{figure}[t]
  \centering
  \begin{subfigure}[b]{\linewidth}
  \includegraphics[width=\linewidth]{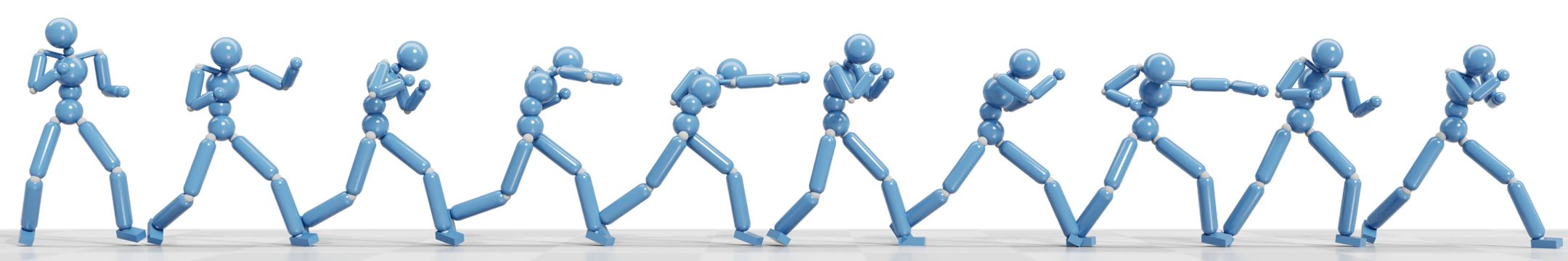}
  \caption{Diverse punch behaviors, given different initial stances, performed by a single policy learned from reference clips in the ``punch'' collection (Table~\ref{tab:motion_collections}). }
  \label{fig:punch_diverse}
  \end{subfigure}
  
  \begin{subfigure}[b]{\linewidth}
  \centering
  \includegraphics[width=\linewidth]{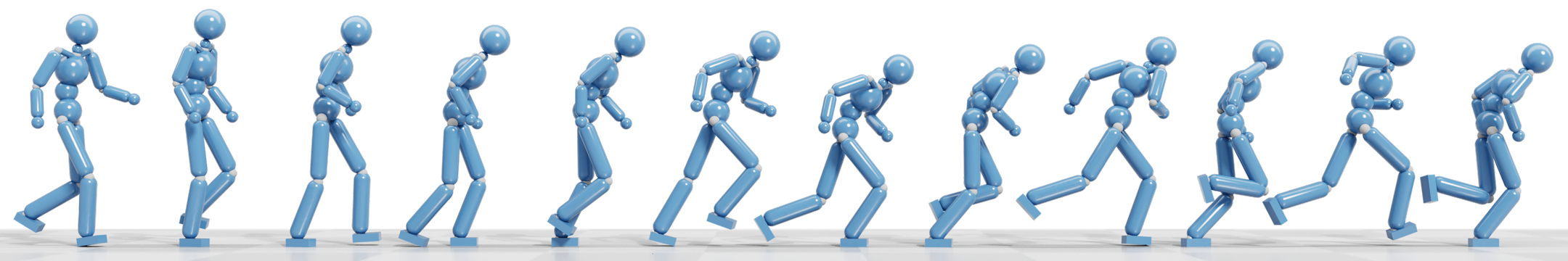}
  \hfill
  \includegraphics[width=\linewidth]{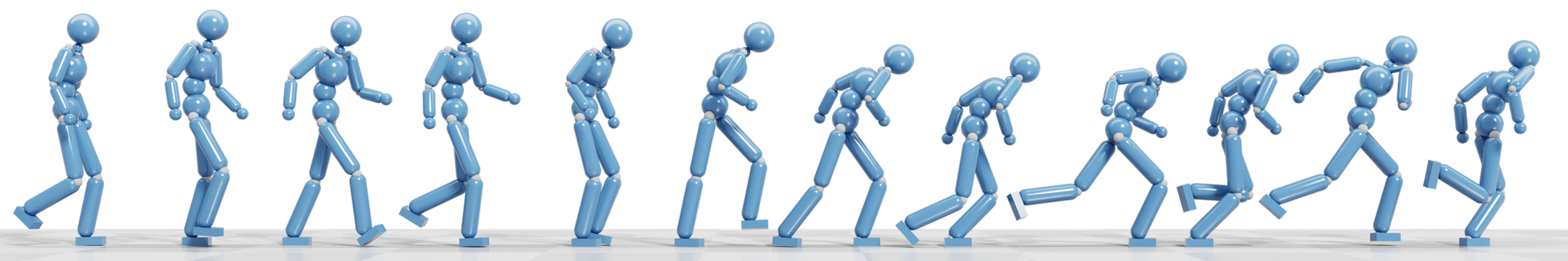}
  \caption{Diverse transitions from walking to running performed by a single policy learned from reference clips in the ``run'' collection (Table~\ref{tab:motion_collections}).}
  \label{fig:run_from_walk_diverse}
  \end{subfigure}
  
  \caption{
Diverse behaviors exhibited by one policy learned from a collection of reference clips in an integrated manner. 
The policy can automatically decide which behavior from the reference clips should be taken based on the given initial pose.}
\end{figure}

Besides single-motion clip imitation, our approach can also train a policy in an integrated manner to imitate motions from multiple clips (Table~\ref{tab:motion_collections}). 
As such, we can control 
the character to execute different behaviors automatically according to the initial pose given to the policy that takes over the character.
For example, one policy can control the character to perform different kinds of punches according to the character's stance, as shown in Fig.~\ref{fig:punch_diverse}. 
In Fig.~\ref{fig:run_from_walk_diverse}, a single policy, learned from two reference clips of running motion starting 
with a different foot on the ground, allows the character to perform flexible walk-to-run transitions.

When the same number of samples is exploited for training,
learning from multiple clips in an integrated manner exploits only a part of those samples to train each distinct behavior.
The difficulty of learning, therefore, would increase if poses from these clips are quite distinct.
For example, the clips used in the ``run'' collection have similar poses during the running phase, and the policy only needs to learn two different transitions besides running (walk to run starting with the left and starting with the right foot on the ground, respectively).
In contrast, motions in the ``spinning and long jump'' collection,
differ a lot to each other, as shown in Fig.~\ref{fig:motions}, and
the imitation error increases compared to the corresponding single-motion clip imitation.
In our experiments, the largest collection for single policy training contains five clips of ``get up' motions with different poses lying on the ground, having a total length of 13.7s (411 frames). 
The trained controllers can imitate different behaviors in all the five clips with a small error, as shown in Table~\ref{tab:motion_collections}.

\subsection{Robustness}
We evaluate the policy robustness through projectile testing. 
During the test, at each frame, a cube 
with fixed mass is launched  towards a uniformly sampled location on the torso of the character, which is controlled by the ``walk'' policy. 
All cubes have a side length of 0.2m and an initial velocity of 0.2m/s. 
We consider a control policy to fail if the character falls down and keeps lying on the
\begin{wrapfigure}{r}{0.28\textwidth}
  \includegraphics[width=0.28\textwidth]{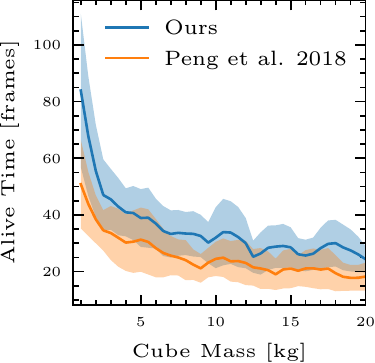}
  \caption{Policy robustness test.}
  \label{fig:robust}
\end{wrapfigure}
ground for 10 consecutive frames. 
For comparisons, we employ a policy trained by motion tracking using the approach from~\citet{peng2018deepmimic} as a baseline. 
Figure~\ref{fig:robust} reports the time, in the term of the number of frames, that a policy can control the character before failure while varying the mass of the cube. 
Solid lines report the average performance over 1,000 trials and shaded regions indicate the standard deviation. 
All policies are obtained without projectile training. 
By controlling the character dynamically based on the last 
pose trajectory,
our approach achieves better robustness performance compared to the baseline algorithm, which keeps tracking the reference motion with a fixed time interval between frames without consideration of the current pose of the character. 

\begin{figure*}[t]
    \centering
    \includegraphics[width=\linewidth]{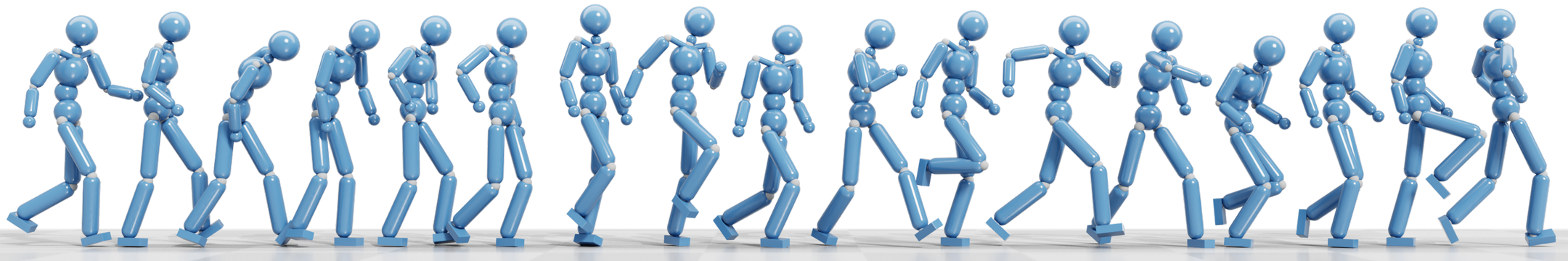}
    
    \vspace{.2cm}

    \includegraphics[width=\linewidth]{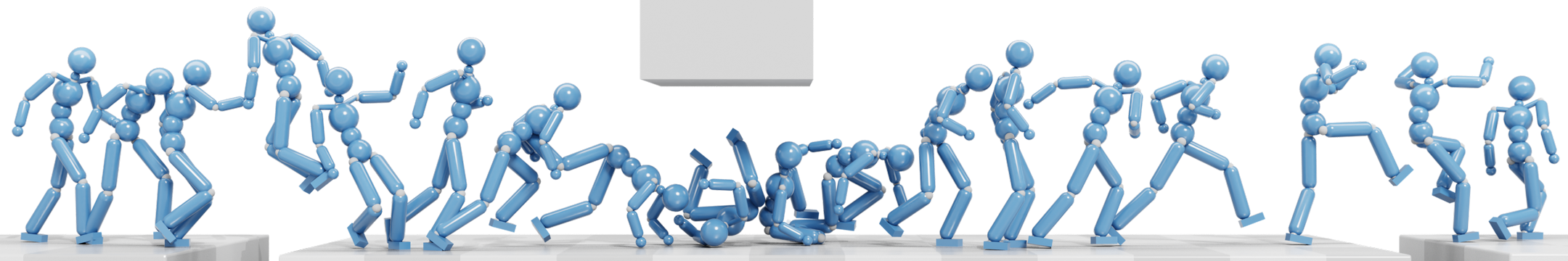}
    
    \vspace{.2cm}
    
    \includegraphics[width=\linewidth]{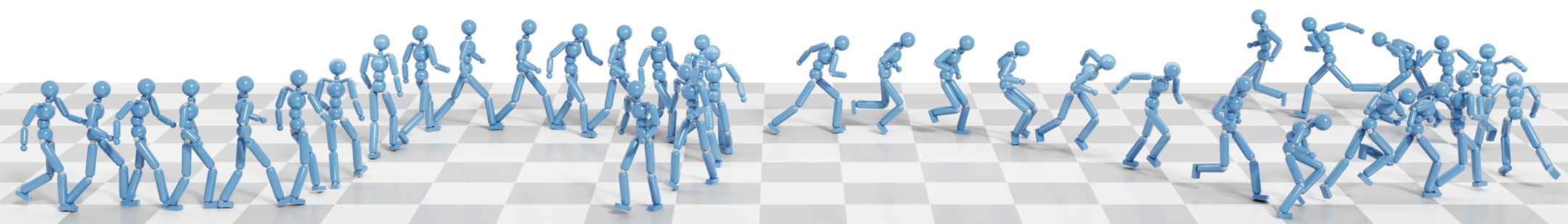}
    \caption{
        Interactive control of a physically simulated character using our approach.
        Responding to the user's input, the character (from top to bottom) walks with varying styles, navigates through a challenging terrain, 
        and changes heading direction.
    }
    \label{fig:policy_switch}
\end{figure*}
\subsection{Interactive Control}\label{sec:int_control}
Our approach allows for interactive character control by taking user control input and performing policy switch based on the average score provided by the ensemble discriminator of the target policy,
as described in Section~\ref{sec:policy_switch}.
Figure~\ref{fig:policy_switch} shows examples of policy switch interactively responding to external control signal provided by the user. 
Please see the supplemental video for more examples.

\begin{figure*}[t]
    \includegraphics[width=0.12\linewidth]{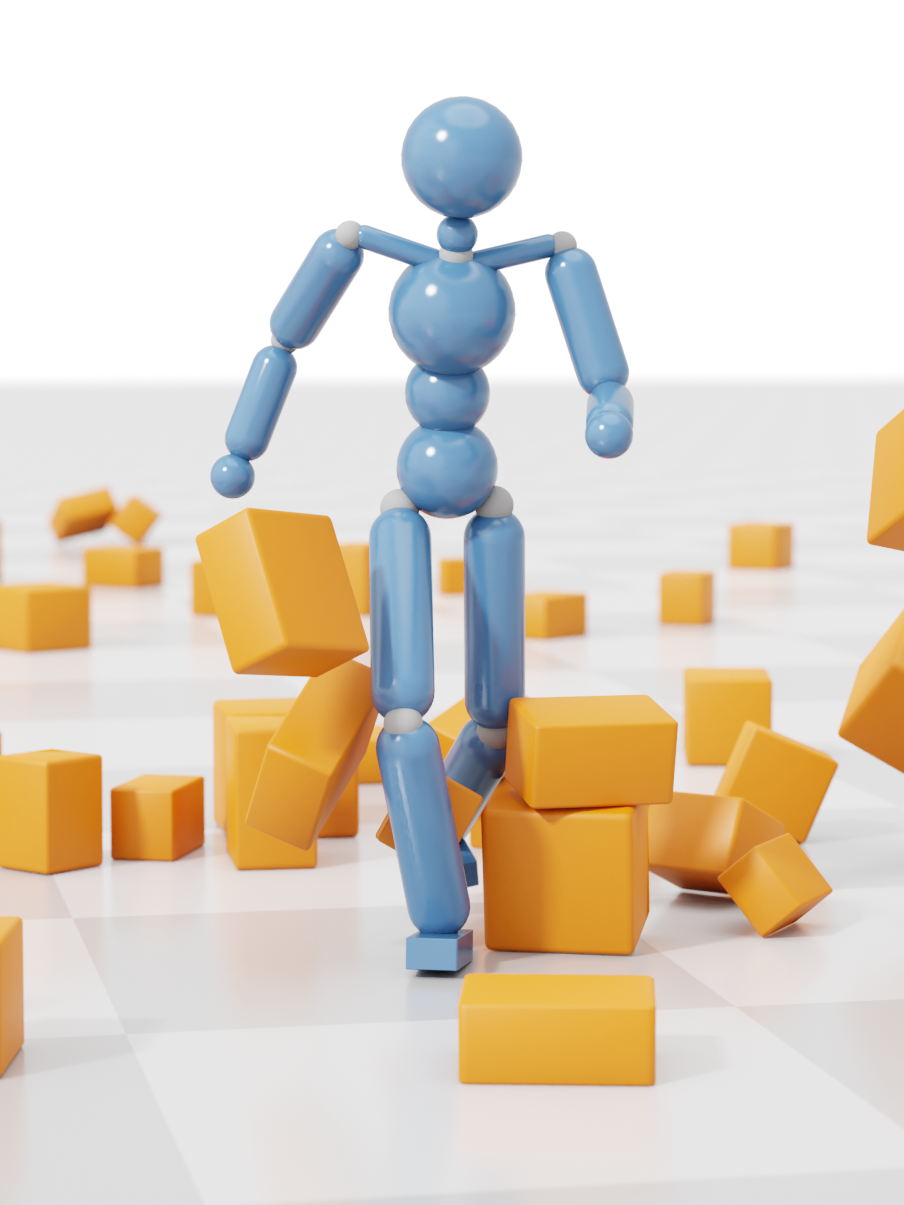}\hfill
    \includegraphics[width=0.12\linewidth]{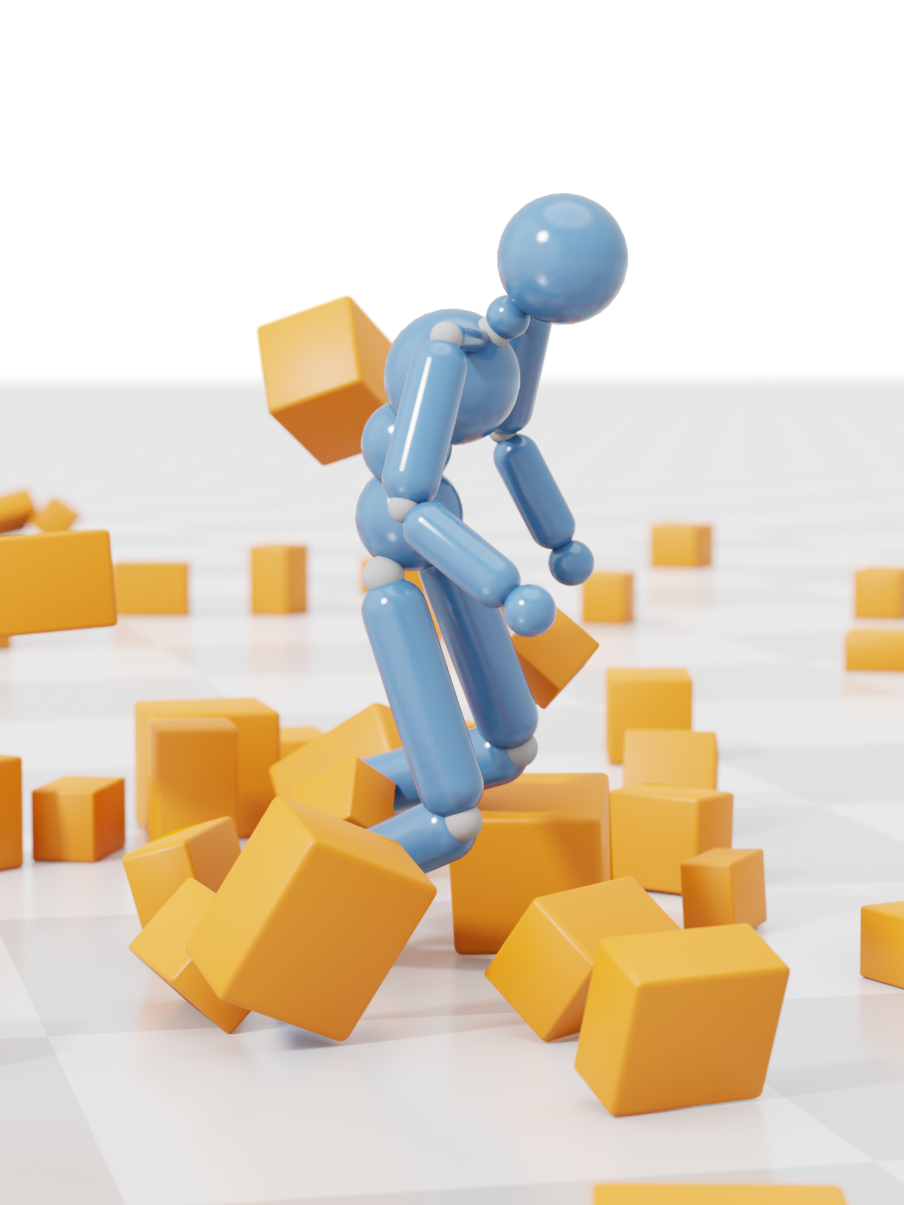}\hfill
    \includegraphics[width=0.12\linewidth]{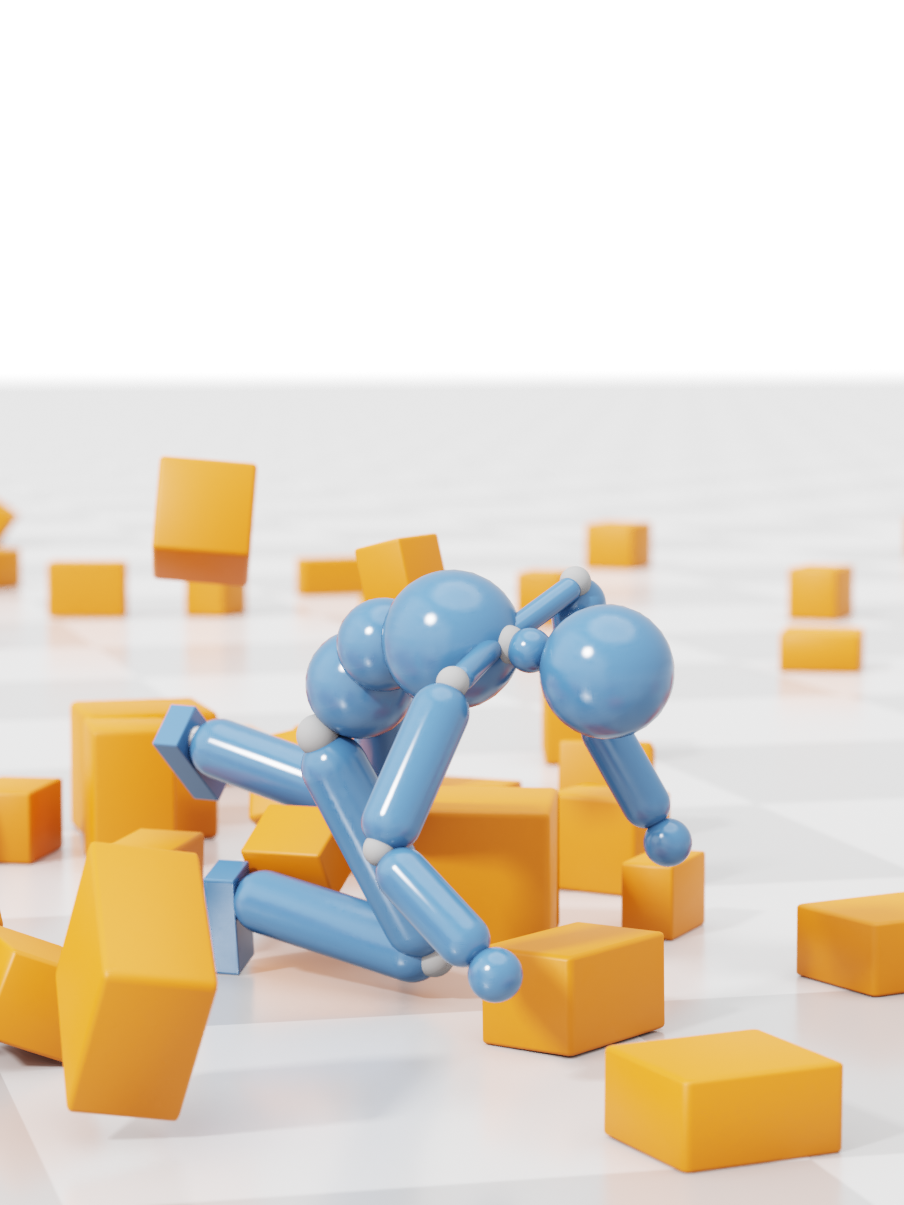}\hfill
    \includegraphics[width=0.12\linewidth]{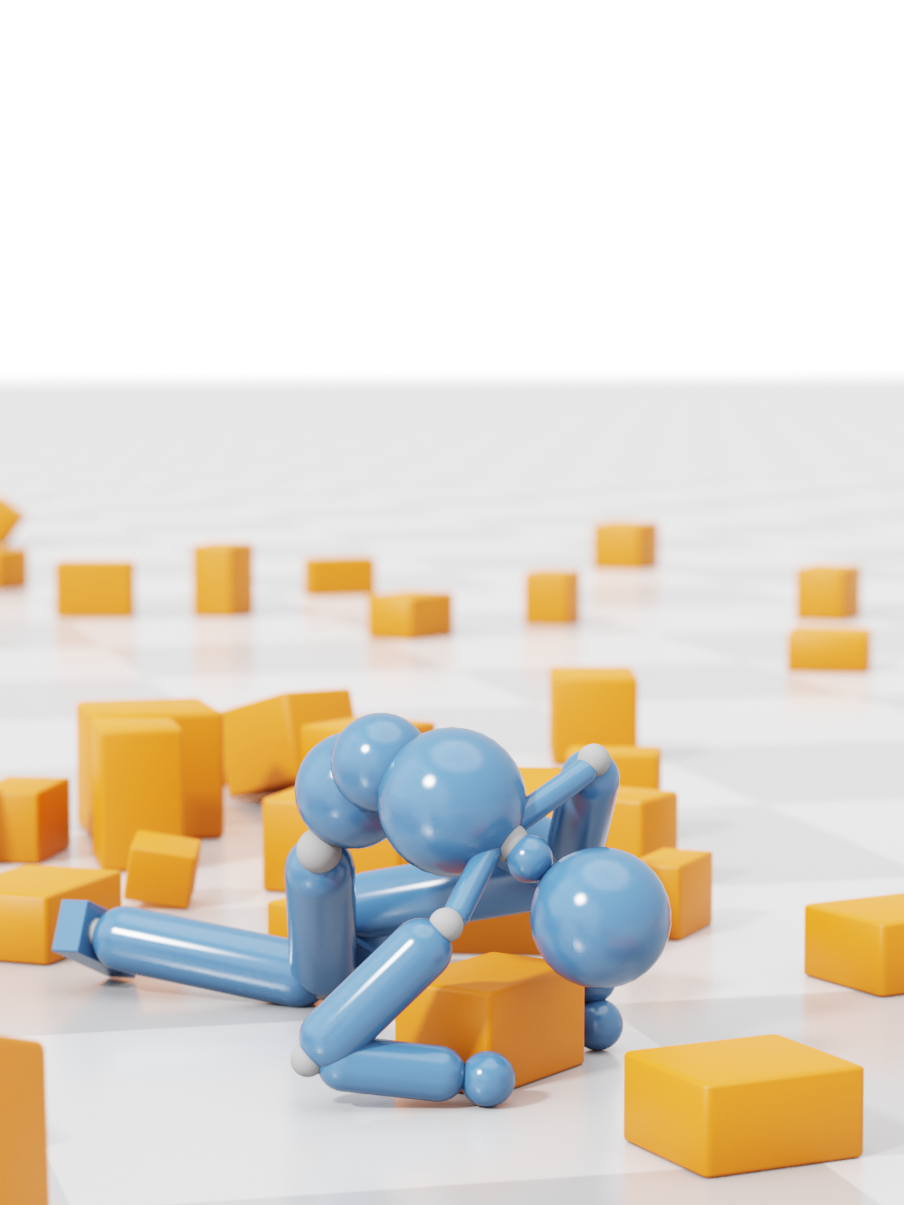}\hfill
    \includegraphics[width=0.12\linewidth]{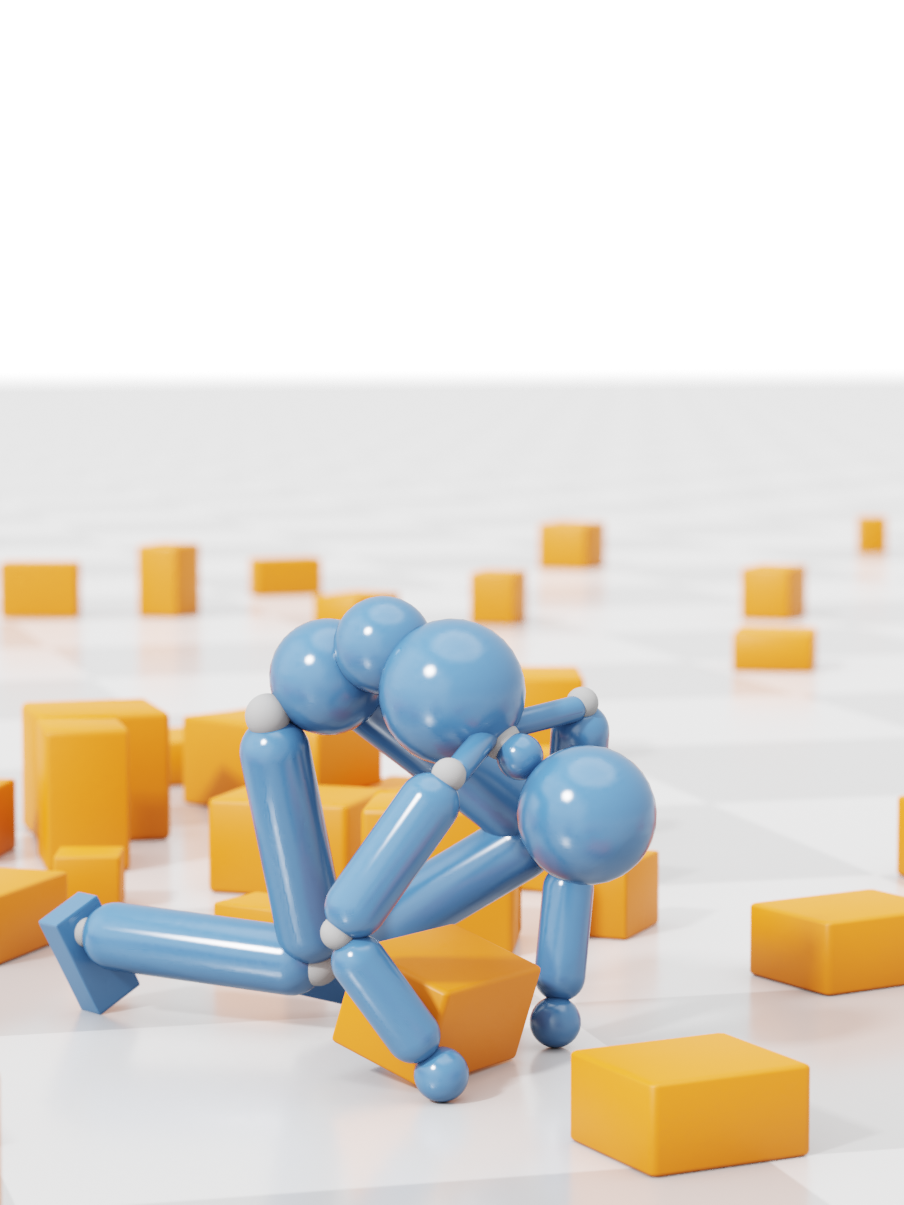}\hfill
    \includegraphics[width=0.12\linewidth]{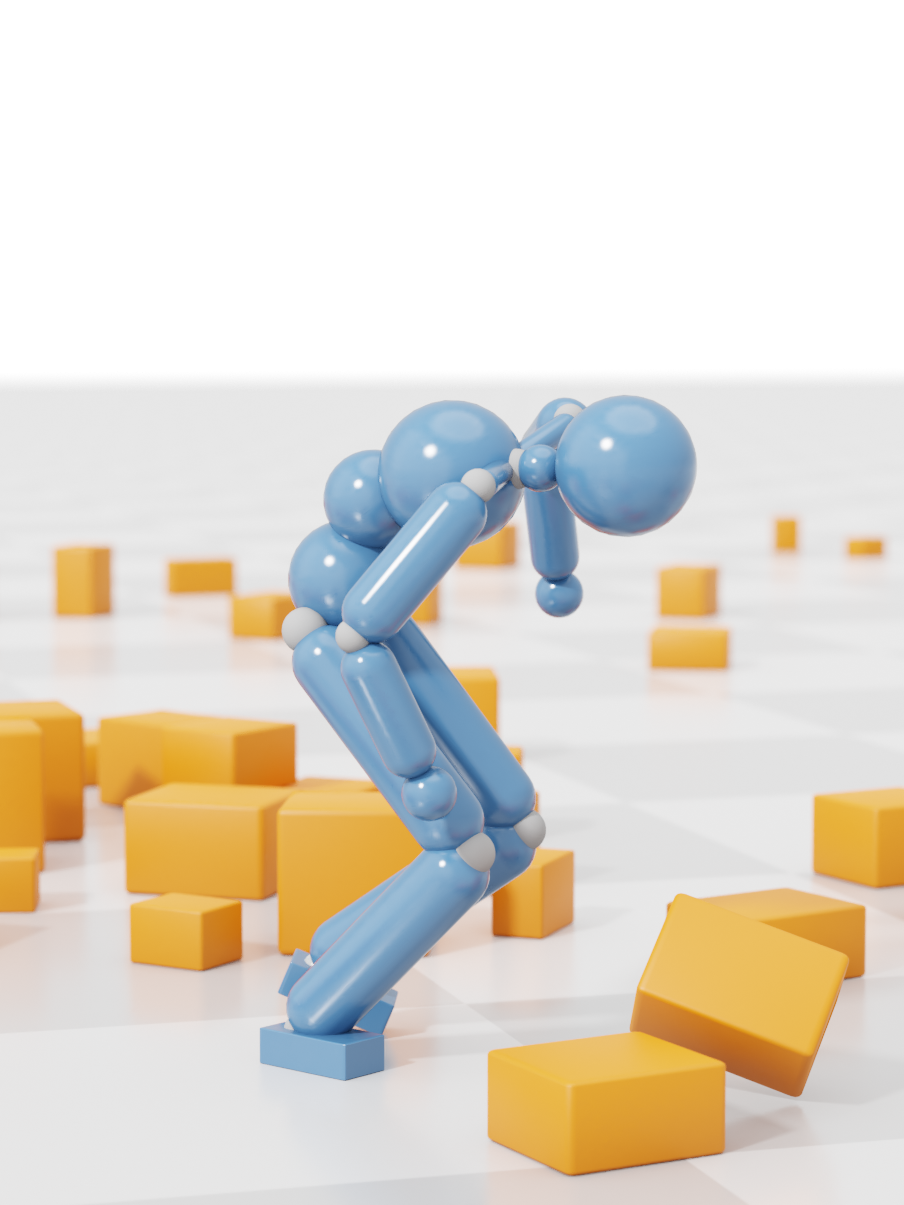}\hfill
    \includegraphics[width=0.12\linewidth]{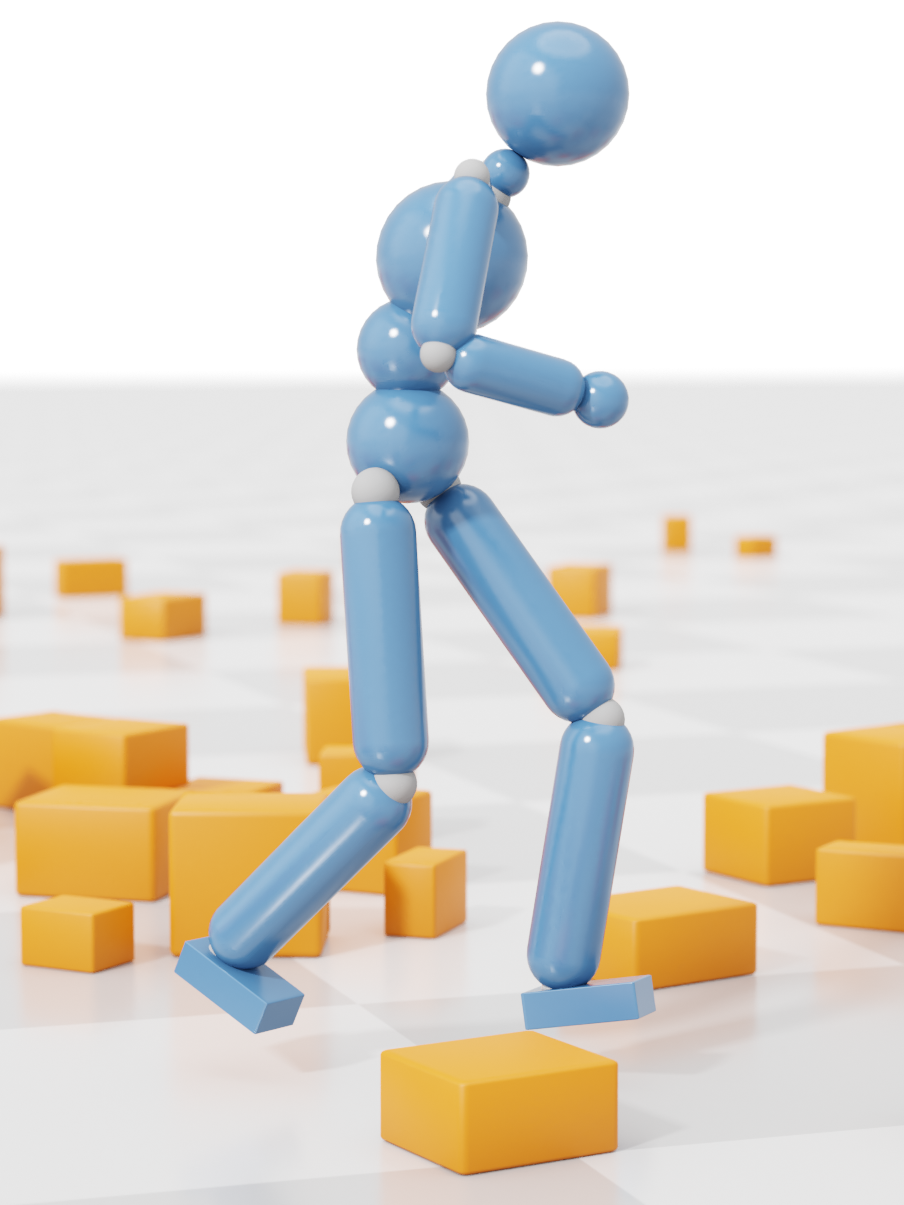}\hfill
    \includegraphics[width=0.12\linewidth]{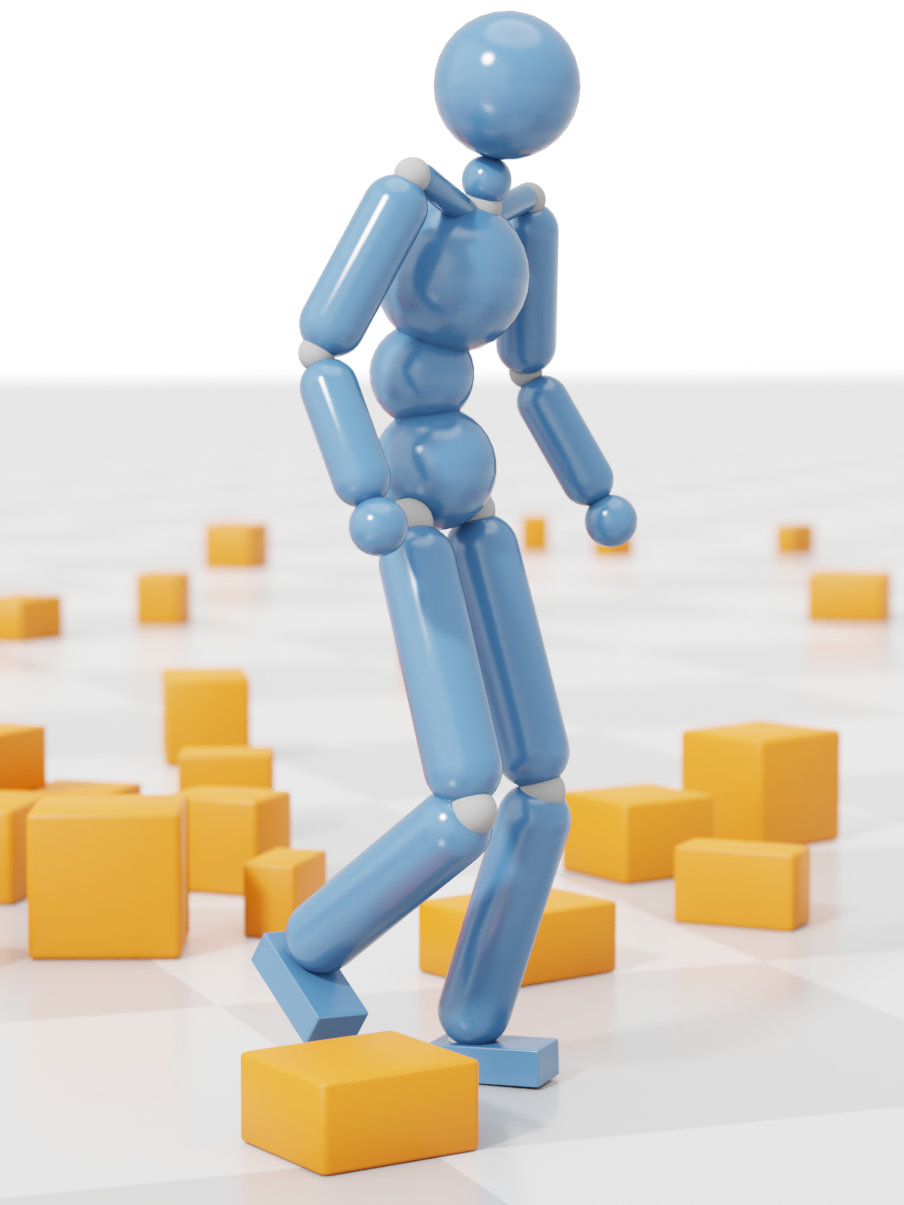}
  \caption{Auto-activated recovery policy.
  A get-up policy works as a recovery policy activated automatically to help a fallen down character to get back to walking. 
  The get-up policy is obtained by learning from reference clips in the ``get up'' collection integratedly (Table~\ref{tab:motion_collections}).} 
  \label{fig:recovery}
\end{figure*}

Besides being a system passively responding to user control input, our interactive control scheme can also work actively by setting up a recovery policy to check and, if necessary, temporally take over the character.
Figure~\ref{fig:recovery} shows a get-up policy working as a recovery policy to help a fallen down character get back to walking.
The get-up policy keeps checking the character state by querying the score from its discriminators at each frame.
It is activated automatically when the score is greater than a threshold value,
and returns the control back to the ``walk" policy if the ensemble discriminator from the ``walk'' policy gives a score high enough.

\subsubsection*{Transition Test}
\begin{figure}[t]
    \centering
    \begin{subfigure}[b]{0.49\linewidth}
    \includegraphics[width=\linewidth]{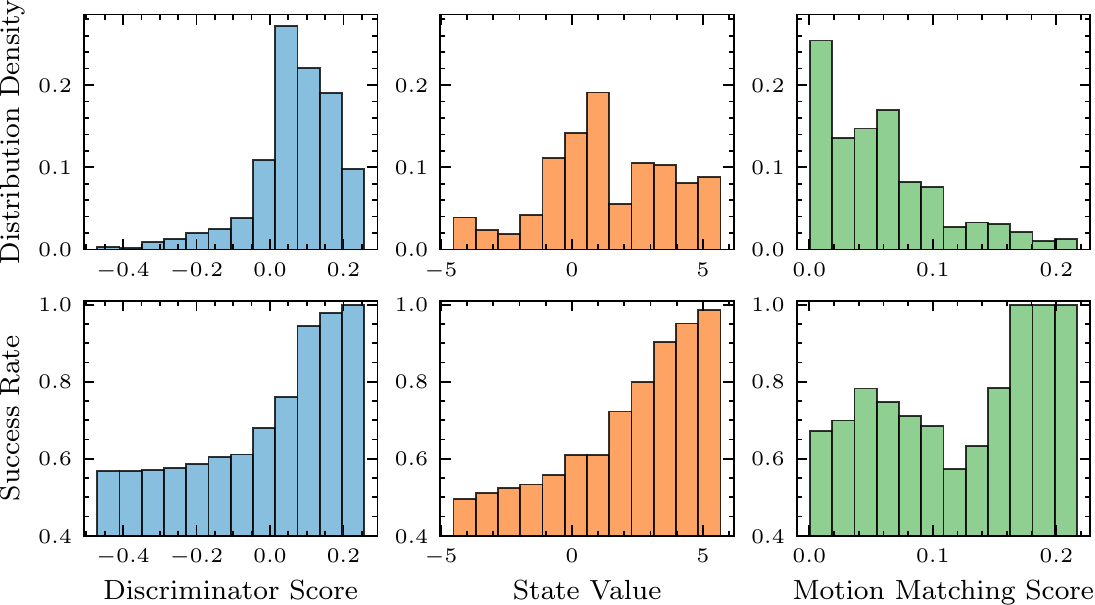}
    \caption{Switch from ``long jump'' to ``run''.
    }
    \label{fig:resp_test_jump_run}
    \end{subfigure}\hfill
    \begin{subfigure}[b]{0.49\linewidth}
    \includegraphics[width=\linewidth]{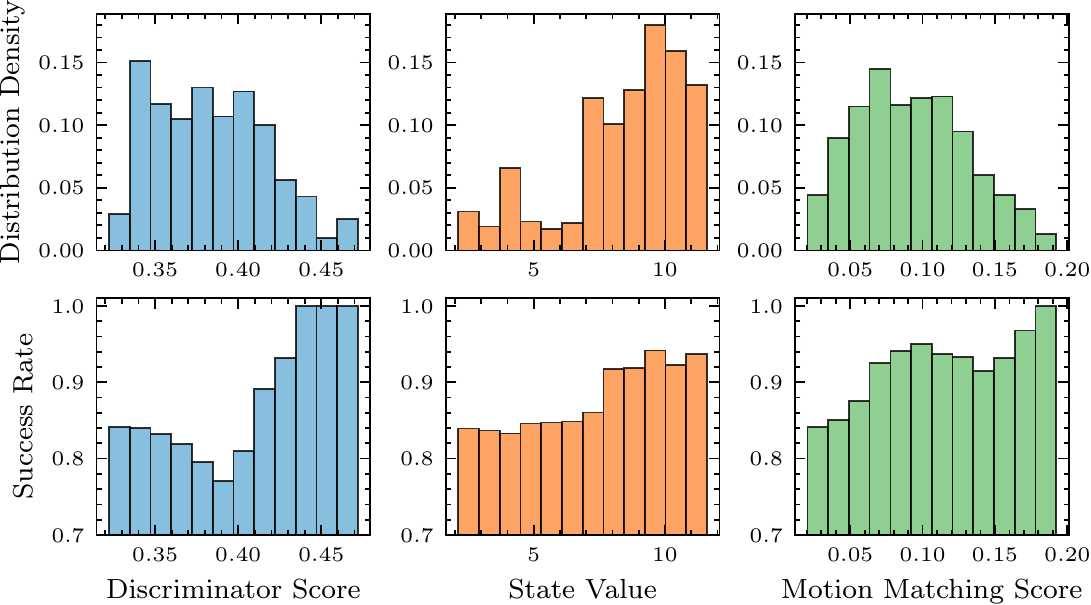}
    \caption{Switch from ``run'' to ``long jump''.
    }
    \label{fig:resp_test_run_jump}
    \end{subfigure}
    
    \caption{Evaluating different metrics for policy switch. The top row reports the distribution of scores/values per metric obtained when the target policy evaluates the character poses controlled by the source policy. The bottom row reports the success rate of policy switch for different scores/values used as the switch threshold.
    }
    \label{fig:resp_test}
\end{figure}
To evaluate the performance of the ensemble discriminator score during interactive control, we run 10,000 tests of policy switching  
from the ``long jump'' policy to the ``run'' policy.
In each trial, the character is initially controlled by the ``long jump'' policy with a pose randomly sampled from the reference clip. 
We consider the switch to be successful if the ``run'' policy can take over during the long jump motion allowing the character to safely perform running.
Figure~\ref{fig:resp_test_jump_run} shows the corresponding score distribution 
and provides statistics on the success rate when different discriminator scores are used as a decision threshold to perform policy switch. 
We also report the results of policy switch from ``run'' to ``long jump'' in Fig.~\ref{fig:resp_test_run_jump}. 
As shown in the figures, when facing unseen poses from the other motion, the discriminator ensemble still provides a reliable measurement to reflect the success rate of policy switch.

As a comparison, we employed the value network in our DPPO implementation and used the state value to perform policy switch check, with the corresponding results shown in Fig.~\ref{fig:resp_test}.  
The state value provides an estimation of the accumulated reward, which is actually the discriminator score (Eq.~\ref{eq:reward} and~\ref{eq:switch_cond}), and can reflect the policy switch success rate. Our implementation, though, directly uses the discriminator score to perform policy switch check,
as we find that a higher state value cannot always guarantee a higher success rate, for example, during the switch from the ``run'' policy to ``long jump''.
In Fig.~\ref{fig:resp_test_jump_run}, we also report the policy switch performance when using the inverse of motion matching cost~\cite{clavet2016motion} as a similarity score. 
The issue with this approach is that the sample coverage rate decreases too much as the chosen threshold value increases.
This means that when a high motion matching score is chosen as the decision threshold to guarantee successful transitions, a large portion of valid policy switches could be denied.
We refer to the supplementary document for more results about the success rate of policy switch in various scenarios.

\begin{figure*}[t]
    \centering
    \begin{subfigure}[b]{\linewidth}
    \centering
    \includegraphics[width=\linewidth]{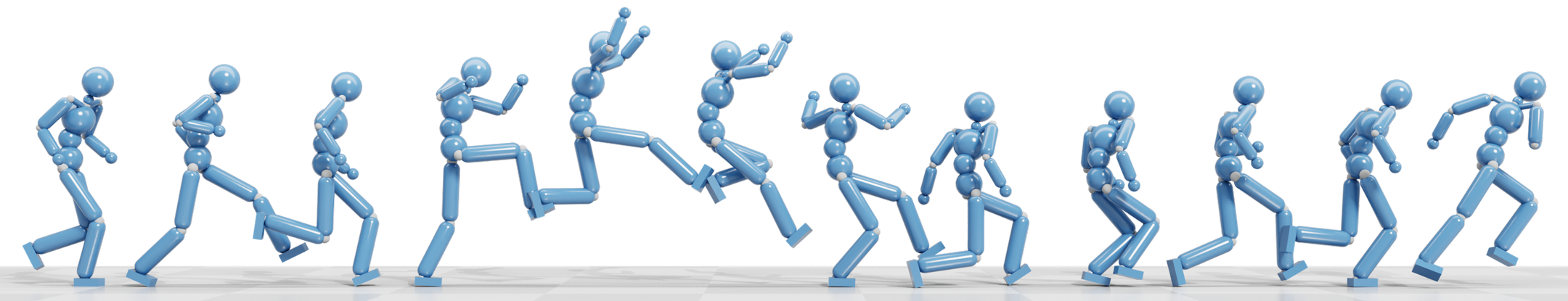}
    \caption{Correct implementation of policy switch of running-jumping-running.}
    \label{fig:succ_switch}
    \end{subfigure}
    
    \begin{subfigure}[t]{\linewidth}
    \centering
    \includegraphics[width=\linewidth]{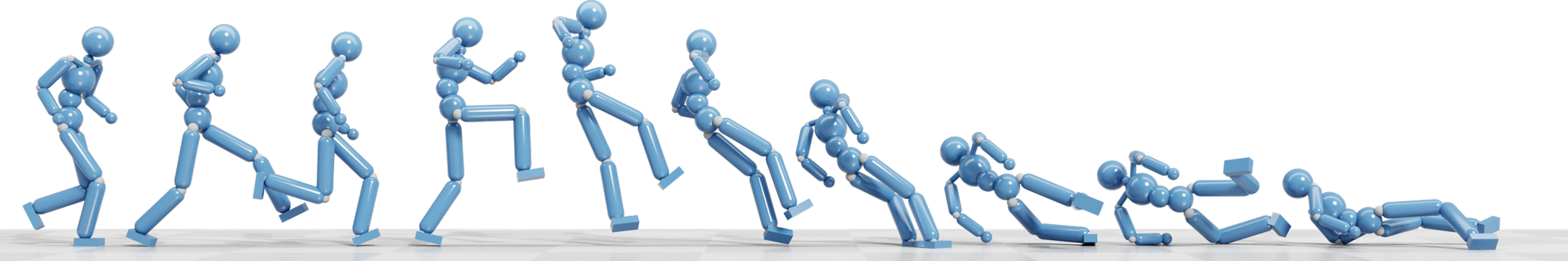}
    \caption{Forcefully switching policy from jumping to running, though discriminators give a low score, and character falls down.}
    \label{fig:fail_down}
    \end{subfigure}
    
    \begin{subfigure}[t]{\linewidth}
    \centering
    \includegraphics[width=\linewidth]{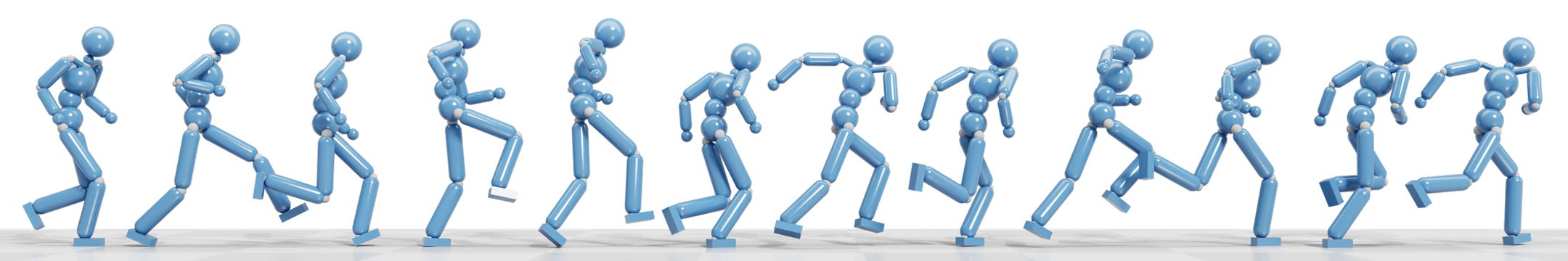}
    \caption{Control switches back to the running policy too early before the character makes a jumping action.}
    \label{fig:fail_delay}
    \end{subfigure}
  \caption{Failure case study of policy switch between running and jumping.
}
  \label{fig:fail}
\end{figure*}
\subsubsection*{Failure Case Study}
Policy switch typically fails due to two reasons. The first is that the target policy fails to control the character given its current pose.
The character under such a case would fall down and be unable to recover, as, e.g., shown in Fig.~\ref{fig:fail_down} where we force a policy switch to happen with a low score value.
This issue can be solved by choosing a proper threshold value to indicate if the switch is possible, which can be behavior specific. 
For example, as shown in Fig.~\ref{fig:resp_test}, 
a threshold value of the discriminator score of 0.1 can be chosen to ensure a success rate over 95\% when switching policy from jumping to running;
and to guarantee a safer switch, the threshold value can be chosen as 0.2.
The second kind of failure is that 
a target policy takes over the character before the current policy finishes its expected action.
This is not a real issue, if the target policy simply takes over as fast as possible responding to the user's control input.
However, it could be a problem when we need the policy switch to occur \emph{after} a specific action done.
For example, as shown in Fig.~\ref{fig:succ_switch}, 
after the character performs a jump,
we would like to automatically switch the control policy back to running.
If we query the discriminators of running policy right after the jumping policy takes over, 
the discriminators will give a higher score, 
since the character is still in the pose of running;
and, hence, the control policy will switch back to running without the character performing any jump action, 
Figure~\ref{fig:fail_delay} shows an example of such a failure.  
This issue can be addressed by introducing frame delays between policy switches.
For example, given that the long jump motion has 43 frames in total, after its corresponding policy takes over the character,
we can set a delay of 20 frames before the system calls on other policies.
As such, the long jump policy has enough time to let the character perform a jump before any other policy takes over.

\subsection{Runtime Cost}
\begin{wrapfigure}[8]{r}{0.48\textwidth}
\footnotesize
    \centering
    \begin{tabular}{lcc}
    \toprule
        & Policy Network & Discriminators \\
    \midrule
    \# of Parameters & 1,154,730 & 381,810 \\
    Model Size       & 4.62 MB    & 1.53 MB  \\
    Forward Runtime  & 189.52 $\pm$ 8.71$\mu$s & 193.53 $\pm$ 5.61$\mu$s\\
    \bottomrule
    \end{tabular}
    \captionof{table}{Runtime cost of single policy and discriminator ensemble networks.}
    \label{tab:runtime}
\end{wrapfigure}
We test models trained using our approach on a common consumer-level computer with Intel i7-7700K CPU and GeForce 1080Ti GPU and report the runtime cost in Table~\ref{tab:runtime}. 
The listed number of parameters and model size are for a single control policy and an ensemble of 32 discriminators. 
The forward runtime reports the average time over 1,000 trials that it takes to perform a forward pass on the policy network and the discriminator network, respectively, of each of the five controllers
shown in Table~\ref{tab:motion_collections}.  
As described in Section~\ref{sec:setup}, our approach uses a fixed network architecture for each controller, and thus each policy and the corresponding ensemble discriminator have a fixed model size, which is unrelated to training data set, i.e., the length and the number of reference clips that a policy is trained to imitate. 
The total memory usage will increase linearly with the number of controllers employed in one scenario.
Even though the ensemble of discriminators has a smaller memory footprint than the policy, it requires a bit more time for a forward pass because the observation space has one extra temporal dimension compared to the state space.

\subsection{Ablation Studies}\label{sec:sensitivity}
\begin{wrapfigure}[16]{r}{0.49\textwidth}
  \vspace{-0.4cm}
  \includegraphics[width=\linewidth]{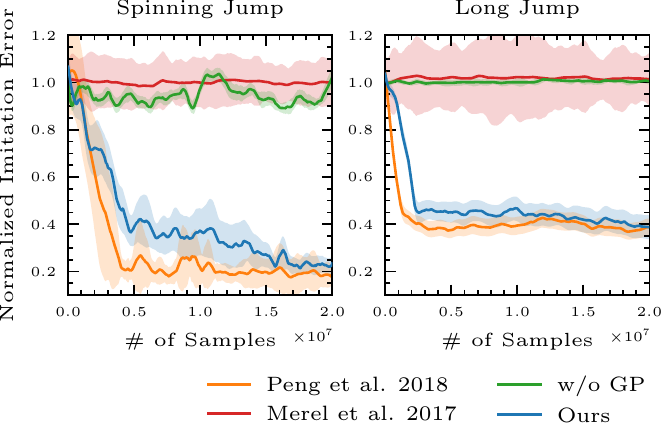}
  \caption{Gradient penalty ablation study. Colored regions denote mean values $\pm$ a standard deviation based on 20 trials.}
  \label{fig:ablation}
\end{wrapfigure}
We perform ablation studies on various design decisions of our
GAN-like approach using two challenging tasks: spinning jump and long jump.
Spinning jump requires the character to keep balance well in the air and spin its body correctly.
The long jump task has a demand on the jump distance and height, as the imitation error is measured by the global positions of body links.
As shown in Table~\ref{tab:motions}, 
the trained policies perform relatively worse in these two tasks, compared to the policies for other control tasks. 

Figure~\ref{fig:ablation} focuses on the gradient penalty loss term for discriminator training, and compares the imitation error during training to two baselines:
the work from~\citet{peng2018deepmimic} that relies on motion tracking and the work from~\citet{merel2017learning} that uses GAIL with observations as the input to the discriminator. 
As it can been seen, 
our approach can achieve comparable performance to the state-of-the-art baseline from~\citet{peng2018deepmimic} that has a carefully designed reward function to facilitate imitation learning. 
Training a physics-based humanoid character controller using the method from~\citet{merel2017learning} typically needs hundreds of millions of samples,
and cannot work effectively within 20 million of samples. 
When no gradient penalty is used with our approach (``w/o GP''), the discriminators can easily identify samples from the character agent. Hence, the policy network always receives a very low reward signal and hardly learns anything useful.

\begin{figure}[t]
    \centering
    \begin{subfigure}[b]{0.48\linewidth}
    \includegraphics[width=\linewidth]{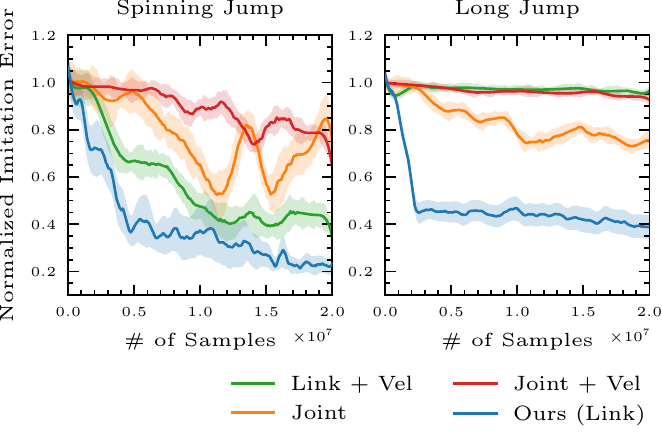}
    \caption{Ablation study on observation spaces.}
    \label{fig:sen_sspace}
    \end{subfigure}\hfill
    \begin{subfigure}[b]{0.48\linewidth}
    \includegraphics[width=\linewidth]{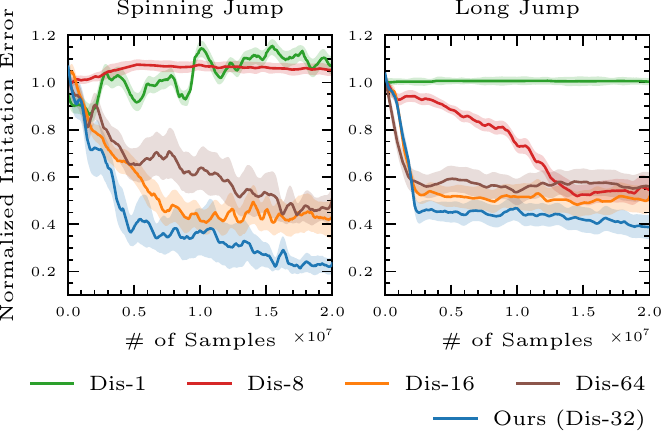}
    \caption{Ablation study on number of discriminators used.}
    \label{fig:sen_dout}
    \end{subfigure}
  \caption{
  Effect that different observation spaces and number of discriminators have on the learning performance.
  (a) Link + Vel uses the link velocity besides just the position and orientation as the observation space for the discriminators. 
  The Joint space uses the local joint pose plus the root position and orientation, with the Joint + Vel also including the joint and root velocity. 
  (b) Dis-N denotes the case of training N discriminators simultaneously for reward generation.
  } 
  \label{fig:sens}
\end{figure}

Figure~\ref{fig:sen_sspace} shows the performance when different observation spaces are used for discriminator training.
By default, our approach uses as the observation space the position and orientation of character body links from the last 4 consecutive frames, in the local system defined by the root position and orientation of the last frame.
As it can been seen, the introduction of velocity either from 
links (Link + Vel) or joints (Joint + Vel) could increase the difficulty of training.
It is relatively hard for the policy to control the immediate link or joint velocity through a PD servo in order to meet the reference velocity computed from the pose difference between two keyframes in the reference motion.
Therefore, discriminators can distinguish character pose trajectories from the reference motion easily based on velocity terms,
and generate quite low reward signals, even though the pose trajectories are visually similar.
As for the visual performance, the introduction of velocity is redundant, since the visual velocity can be inferred from character poses in consecutive frames.
The observation space that employs local joint poses (Joint) provides worse performance compared to our approach that uses the link positions. 
Discriminators in the local joint case could be fooled by the policy network and cannot distinguish samples from the reference motion effectively. 
This is due to the fact that the error between joint poses measured by rotation angle for revolute joints and quaternion for spherical joints is typically small, even though the overall character poses are visually quite different. 

In Fig.~\ref{fig:sen_dout}, we compare the training performance when different number of discriminators are employed.
As shown in the figure, 
the policy network with 1 discriminator and that with 8 discriminators cannot be trained effectively. 
The more discriminators are used, the harder the ensemble discriminator training will be. 
This avoids overfitting during the early stage of training where the character agent performs largely differently to the reference motion, and helps solve the problem of vanishing reward signal. 
On the other hand, the increased difficulty in discriminator optimization prevents the discriminators from effectively identifying small errors between the character poses and the reference motion, as shown in the 64 discriminators case. 
We use an ensemble of 32 discriminators to balance the trade-off between overfitting and underfitting.

\section{Discussion and Future Work}\label{sec:conclusion}
We propose a GAN-inspired approach combined with reinforcement learning to perform imitation learning for physics-based character control without requiring any explicit design of a reward function.
In our approach, policies perform inference and control the character through short-term pose trajectories, 
eliminating the need to indicate the target keypose that it should follow.
Based on this approach, by exploiting the discriminator to measure the similarity between the character's pose and the reference motion, we implement an interactive control system for policy switch during character control according to external user-input signals. 
Although, we empirically demonstrated that our approach can perform imitation learning and interactive control effectively,
there are still some limitations that we would like to address in the future.

\paragraph{Motion transition quality}
In our experiments, the reference clips are extracted from the LAFAN1 dataset (Section~\ref{sec:data_acq}) and are typically discontinuous to each other.
Each control policy only learns to imitate behaviors from its own collection of reference clips,
and is not fine-tuned to adapt to any pose in other clips.
Consequently, this could often lead to nonsmooth transitions during policy switch, or even make an expected transition unfeasible because there are no similar poses in the reference motions of the two policies.
In practice, this issue can be addressed by 
explicitly providing the transition motions, which can be obtained from motion capture or in-between motion generation techniques~\cite{harvey2020robust}, and using them as reference motions for the controller policies to imitate. 
An alternative solution is to force a transition stage in training,  during which the character is initialized using poses from other motions~\cite{bergamin2019drecon}.

\paragraph{Unweighted observations.} 
In our framework, discriminators distinguish character pose samples from the reference motion via root-related link positions and orientations,
and rely on these observations to provide reward signals for policy training.
A manually designed reward function for imitation learning~\cite{peng2018deepmimic,bergamin2019drecon,ScaDiver} would typically consider the contributions that different body links or joints make for specific behaviors,
and assign different weights to those link- or joint-related errors when computing the reward. 
However, in our current implementation, the observations are unweighted. 
This could cause issues when the policy gives a higher priority during optimization to attenuate the pose error related to some trivial body links or joints, 
which have less impact on visible differences between target poses. 

\paragraph{Long-term memory.} 
As the policy network relies on the poses from the last four frames to perform inference and character control, it lacks the ability of long-term memory. 
When facing complex motions or learning from multiple motion clips, the policy network may fail to learn the diversity of motions if there are some fairly similar short-term pose trajectories in the reference motions leading to different following poses. 
An obvious solution would be to increase the number of poses used to construct the state space. 
However, this could increase the difficulty of learning
as the inference would be disturbed by more noises in the trajectory. 

\paragraph{Importance sampling.}
Our approach can learn diverse behaviors from multiple motion clips. 
In each training episode, 
we initialize the character randomly by uniformly picking a pose from those clips. 
A more effective way of training should take into account the current performance of the policy, 
and do an importance sampling when initializing the character pose.
In such a way, the policy under optimization can dedicate more training to the behaviors that it performed relatively worse, and thus exploit samples more efficiently. 
\citet{park2019learning} use an adaptive sampling method, dividing the reference motion into $n$ segments and initializing the character pose by drawing samples from the segments based on the distribution of state value.
In our approach, we can employ the discriminator score as a more accurate estimation of the performance of the policy under training, and do importance sampling. 

\paragraph{Goal-conditioned learning.}
In interactive control experiments (Section~\ref{sec:int_control}), we only consider the case of switching policies according to external control signals. 
To achieve a truly interactive control policy that is responsive to user input, such as  
a walking or running policy with interactive speed or heading direction adaption, 
we can implement goal-conditioned policy learning by introducing a dynamic goal state during training~\cite{peng2017deeploco,bergamin2019drecon,xie2020allsteps}.
A goal-conditioned reward function combining our GAN-like approach can be set up as
\begin{equation}
    r_t = w_{goal}r_{goal}(\mathbf{s}_t, \mathbf{g}_t, \mathbf{a}_t, \mathbf{s}_{t+1}) + w_{GAN}r_{GAN}(\mathbf{s}_t, \mathbf{a}_t, \mathbf{s}_{t+1}),
\end{equation}
where $r_{goal}$ is the reward related to the goal completion having a dynamic goal state $\mathbf{g}_t$, $r_{GAN}$ is the reward from our GAN-like approach (Eq.~\ref{eq:reward}), and $w_{goal}$ and $w_{GAN}$ are coefficients to balance the weights of the two reward components. 
\\

As we highlight above there are a number of improvements that can be made to further increase the performance and general applicability of our proposed work. 
Based on our current results, we feel that there are several possible applications of our framework in production games and interactive applications as 
i) it does not require any explicit reward design to learn control policies, 
ii) it allows for interactive policy switch during character control based on external input signal at a low runtime cost, 
and iii) it is easy to be extended to support more action policies. 
Of course, a preprocesing step is required to gather reference motions. 
However, we can argue that a technique similar to motion matching can be employed before training to quickly determine whether multiple reference trajectories can be combined to learn an interactive control policy that can support different behaviors. 
While future research is also needed to guarantee the motion quality of controller transitions,
we believe that, overall, our proposed framework provides a simple and flexible alternative for imitation learning and real-time control of physically simulated characters.

\bibliographystyle{ACM-Reference-Format}
\bibliography{sample-base}

\newpage
\appendix
\section{Reference Motion Clip Acquisition}
\label{app:data_acq}
\begin{table}[ht]
\footnotesize
\centering
  \begin{tabular}{lcccc}
    \toprule
    Motion Clip & Collection & File & Start & End \\
    \midrule
    walk & & \verb!run2_subject1! & 5758 & 5791 \\
    pace & & \verb!walk1_subject1! & 3748 & 3799 \\
    limp & & \verb!walk1_subject1! & 4515 & 4572 \\
    swaggering walk & & \verb!walk2_subject1! & 2821 & 2853 \\
    sashay walk & & \verb!walk2_subject1! & 2613 & 2645 \\
    jaunty walk & & \verb!walk2_subject1! & 5002 & 5044 \\
    stomp walk & & \verb!walk2_subject1! & 6581 & 6618\\
    stoop walk & & \verb!walk4_subject1! & 986 & 1014 \\
    joyful walk & & \verb!dance1_subject1! & 2018 & 2054 \\
    walk with arms akimbo & & \verb!walk2_subject1! & 6454 & 6520 \\
    sharp turn during running & & \verb!run2_subject1! & 435 & 479 \\
    90-degree turn during walking & & \verb!walk1_subject1! & 3386 & 3448 \\
    roll & & \verb!pushAndStumble1_subject2! & 4679 & 4780 \\
    fight stance & & \verb!fightAndSports1_subject4! & 1096 & 1110 \\
    \hline
    run & \multirow{2}{*}{run} & \verb!run2_subject1! & 5779 & 5850 \\
    run 2 & &  \verb!run2_subject1! & 5537 & 5608 \\
    \hline
    spinning jump & \multirow{4}{*}{spinning and long jump} & \verb!jumps1_subject1! & 6682 & 6772 \\
    spinning jump 2 & & \verb!jumps1_subject1! & 6858 & 6949 \\
    long jump & & \verb!multipleActions1_subject1! & 3234 & 3287 \\
    long jump 2 & & \verb!multipleActions1_subject1! & 3748 & 3800 \\
    \hline
    get up & \multirow{5}{*}{get up} & \verb!fallAndGetUp3_subject1! & 1318 & 1404 \\
    get up 2 & & \verb!fallAndGetUp3_subject1! & 2016 & 2100 \\
    get up 3 & & \verb!fallAndGetUp3_subject1! & 2247 & 2326 \\
    get up 4 & & \verb!fallAndGetUp3_subject1! & 2473 & 2580 \\
    get up 5 & & \verb!fallAndGetUp3_subject1! & 2717 & 2772 \\
    \hline
    punch & \multirow{3}{*}{punch} & \verb!fightAndSports1_subject4! & 1243 & 1295 \\
    punch 2 & & \verb!fightAndSports1_subject4! & 1109 & 1148 \\
    punch 3 & & \verb!fightAndSports1_subject4! & 1317 & 1369 \\
    \hline
    kick & \multirow{4}{*}{kick} & \verb!fightAndSports1_subject4! & 1394 & 1464 \\
    kick 2 & & \verb!fightAndSports1_subject4! & 971 & 1036 \\
    kick 3 & & \verb!fightAndSports1_subject4! & 1037 & 1099 \\
    kick 4 & & \verb!fightAndSports1_subject4! & 1493 & 1549 \\
  \bottomrule
\end{tabular}
    \caption{Reference motion clips acquired from the LAFAN1 dataset.
\texttt{Start} and \texttt{end} denote the start and end frame, respectively, of the corresponding file in the LAFAN1 dataset from which a reference motion clip is obtained.}
    \label{tab:data_acq}
\end{table}

\newpage
\section{Policy Switch Success Rate versus Discriminator Score}\label{app:resp_test}
\begin{figure*}[ht]
  \centering
  
    \begin{subfigure}[b]{0.49\linewidth}
    \includegraphics[width=\linewidth]{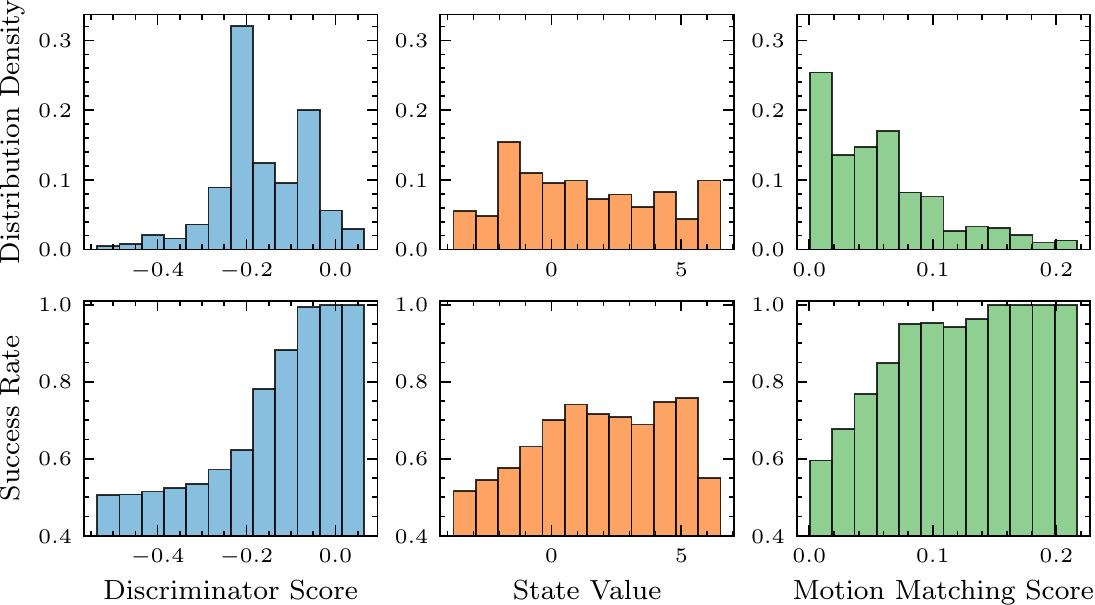}
    \caption{``long jump'' to ``walk''}
    \end{subfigure}\hfill
    \begin{subfigure}[b]{0.49\linewidth}
    \includegraphics[width=\linewidth]{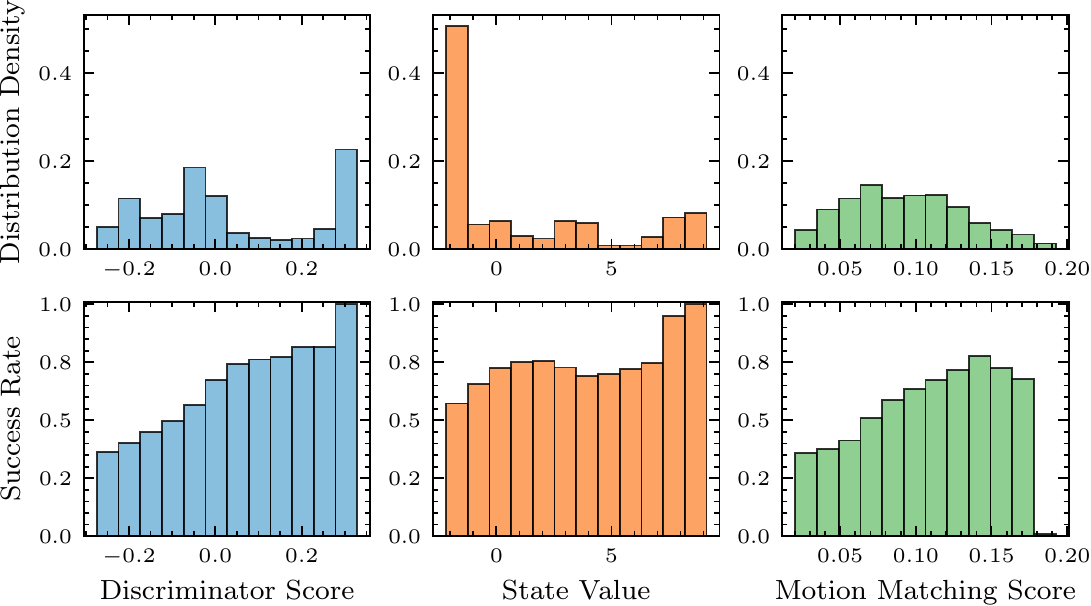}
    \caption{``run'' to ``walk''}
    \end{subfigure}\\\vspace{0.4cm}
    
    \begin{subfigure}[b]{0.49\linewidth}
    \includegraphics[width=\linewidth]{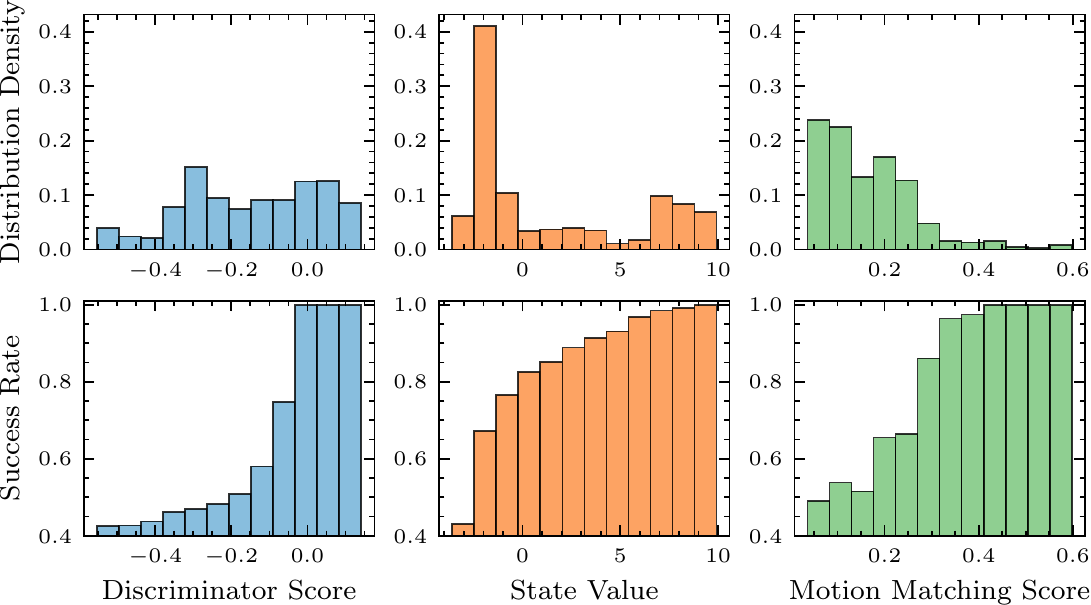}
    \caption{``spinning jump'' to ``walk''}
    \end{subfigure}\hfill
    \begin{subfigure}[b]{0.49\linewidth}
    \includegraphics[width=\linewidth]{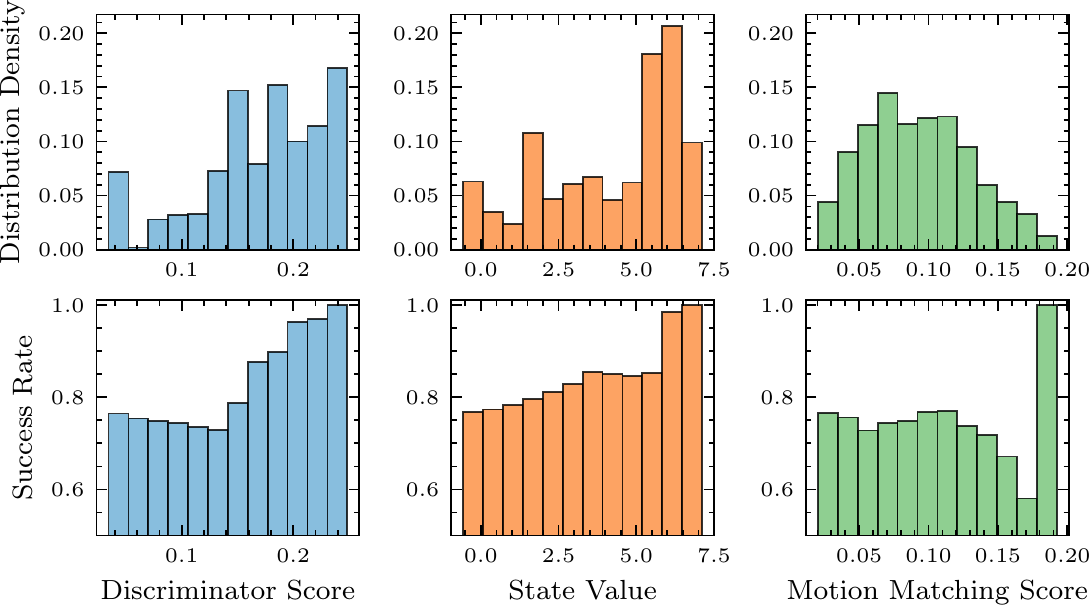}
    \caption{``run'' to ``spinning jump''}
    \end{subfigure}\\\vspace{0.4cm}
    
    \begin{subfigure}[b]{0.49\linewidth}
    \includegraphics[width=\linewidth]{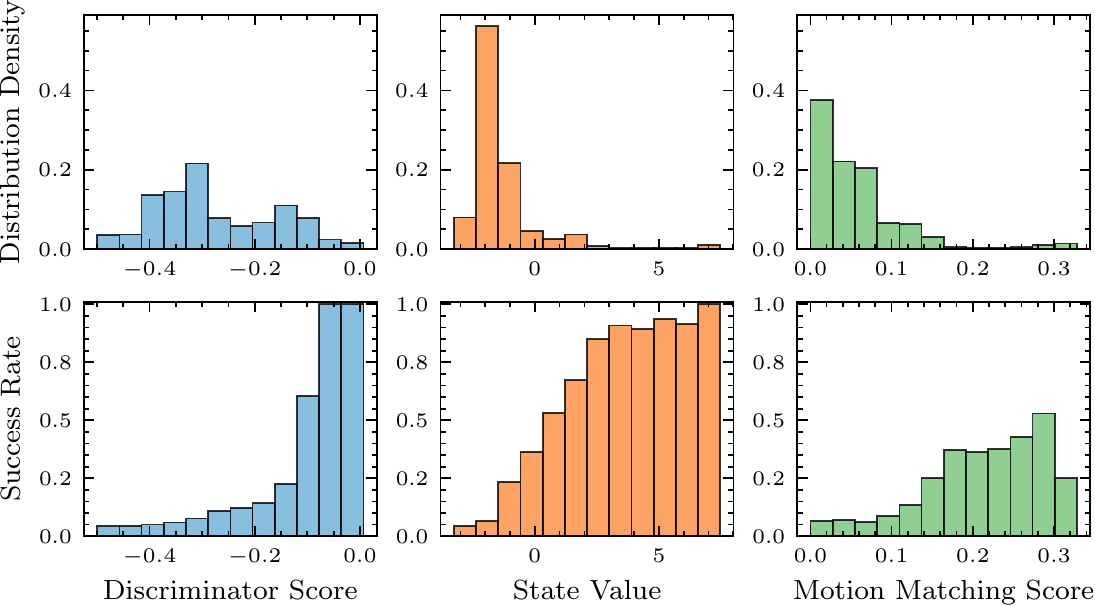}
    \caption{``roll'' to ``walk''}
    \end{subfigure}\hfill
    \begin{subfigure}[b]{0.49\linewidth}
    \includegraphics[width=\linewidth]{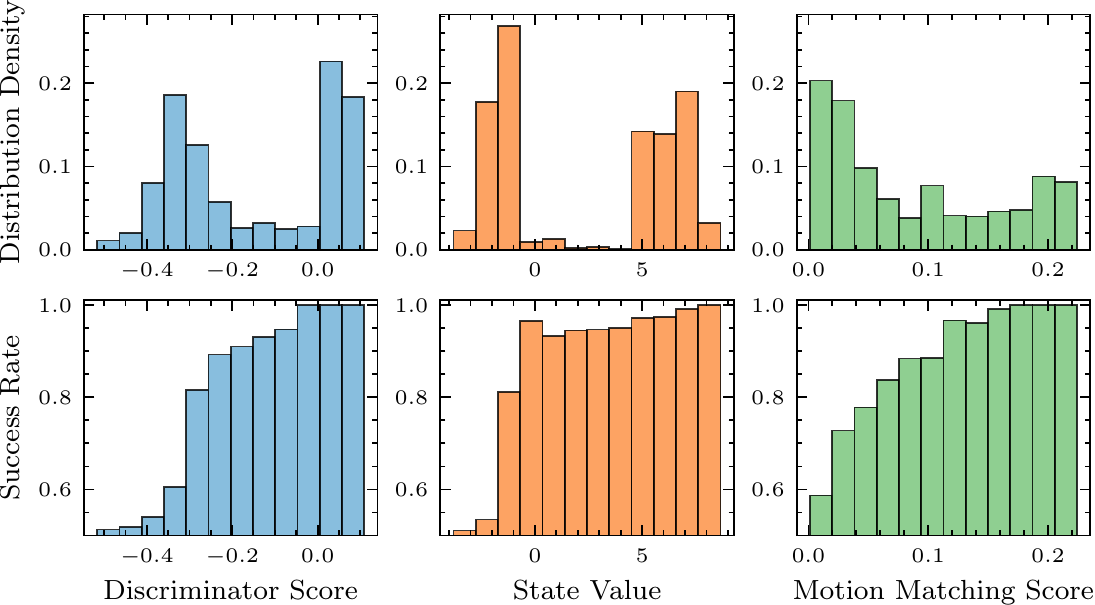}
    \caption{``getup'' to ``walk''}
    \end{subfigure}
    
    \caption{Distributions of different metrics used for policy switch check (top), and the corresponding success rate with different values used as the switch threshold value (bottom). See Section~\ref{sec:int_control} for details of the experiment.}
    \label{fig:resp_test2}
\end{figure*}

\end{document}